\documentclass[12pt]{article}
\usepackage{epsf,rotating,latexsym,astron,amssymb}           
\newcommand{\mc}[3]{\multicolumn{#1}{#2}{#3}}
\begin{document}
\bibliographystyle{astron}

\begin{center}
{\Large \bf Four years of Ulysses dust data: 1996 to 1999}
\end{center}

\bigskip

{\bf
        H.~Kr\"uger$^a$\footnote{{\em Correspondence to:} 
Harald.Krueger@mpi-hd.mpg.de}, 
        E.~Gr\"un$^a$,
        M.~Land\-graf$^b$,
        S.~Dermott$^c$,
        H.~Fechtig$^a$,
        B.~A.~Gus\-taf\-son$^c$,
        D.~P.~Hamilton$^d$, 
        M.~S.~Hanner$^e$,
        M.~Hor\'anyi$^f$,
        J.~Kissel$^g$,
        B.~A.~Lind\-blad$^h$,
        D.~Linkert$^a$,
        G.~Linkert$^a$,
        I.~Mann$^{i,k}$,
        J.~A.~M.~McDon\-nell$^j$,
        G.~E.~Mor\-fill$^g$, 
        C.~Polanskey$^e$,
        G.~Schwehm$^k$,
        R.~Srama$^a$, and 
        H.~A.~Zook$^l$\footnote{Passed away on 14 March 2001}

}
\bigskip

\small
\begin{tabular}{ll}
a)& Max-Planck-Institut f\"ur Kernphysik, 69029 Heidelberg, Germany\\
b)& ESOC, 64293 Darmstadt, Germany \\
c)& University of Florida, Gainesville, FL\,32611, USA \\
d)& University of Maryland, College Park, MD\,20742-2421, USA\\
e)& Jet Propulsion Laboratory, Pasadena, California 91109, USA\\
f)& Laboratory for Atmospheric and Space Physics, Univ.
                 of Colorado, Boulder, \\ 
  & CO\,80309, USA\\
g)& Max-Planck-Institut f\"ur Extraterrestrische Physik, 85748 Garching, 
                                                                   Germany\\ 
h)& Lund Observatory, 221 Lund, Sweden\\
i)&Institut f\"ur Planetologie, Universit\"at M\"unster, 48149 M\"unster, Germany\\
j)& Planetary and Space Science Research Institute, The Open University, \\ 
  & Milton Keynes, MK7 6AA, UK\\
%10)& European Space Research and Technology Center, 2200 AG Noordwijk, 
%                                                      The Netherlands\\
k)& ESTEC, 2200 AG Noordwijk, The Netherlands\\
l)& NASA Johnson Space Center, Houston, Texas 77058, USA\\
\end{tabular}

\normalsize

\bigskip

\begin{abstract}

The Ulysses spacecraft is orbiting the Sun on a highly inclined ellipse ($
i = 79^{\circ}$, perihelion distance 1.3~AU, aphelion distance 5.4~AU). 
Between January 1996 and December 1999 the spacecraft was 
beyond 3~AU from the Sun and crossed the ecliptic plane at aphelion 
in May 1998. In this four-year period 218 dust impacts were
recorded with the dust detector on board. We publish and 
analyse the complete data set of both raw and reduced data for particles with masses
$\rm 10^{-16}\,g$ to $\rm 10^{-8}$~g. Together with 1477 dust impacts 
recorded between launch of Ulysses and the end of 1995 published earlier 
\cite{gruen1995c,krueger1999b}, a data set of 1695 dust impacts 
detected with the Ulysses sensor between October 1990 and December 1999 is 
now available. 
The impact rate measured between 1996 and 1999 was relatively 
constant with about 0.2 impacts per day. The impact direction of the 
majority of the impacts is 
compatible with particles of interstellar origin, the rest are most 
likely interplanetary particles. The observed impact rate is compared with 
a model for the flux of interstellar dust particles. The flux of 
particles several micrometers in size is compared with the 
measurements of the dust instruments on board Pioneer~10 and Pioneer~11
beyond 3~AU (Humes 1980, JGR, 85, 5841--5852, 1980).  Between 3 and 5~AU, 
Pioneer results
predict that Ulysses should have seen five times more ($\rm \sim 10\,\mu\,m$
sized) 
particles than actually detected. 
%Possible explanations are that the
%mass calibrations of Ulysses and Pioneer detectors are incorrect by large
%factors, many Pioneer detections are spurious and not caused by real 
%dust impacts or the Pioneer detections belong to a population of grains 
%not detectable with Ulysses.

\end{abstract}

\section{Introduction}

Ulysses is the first interplanetary spaceprobe which left the ecliptic 
plane of our Solar System and flew over the Sun's poles. The craft's
orbital plane is almost perpendicular to the ecliptic plane with 
an aphelion as far from the Sun as Jupiter. The main 
scientific objectives of the mission are 
the exploration of the Sun at high heliographic latitudes, 
investigation of the solar wind and the interplanetary medium 
as well as interstellar gas and cosmic rays. 

The Ulysses spacecraft is equipped with a highly sensitive
impact ionisation dust detector which measures in situ impacts of micrometre
and sub-micrometre dust grains. The detector is practically identical with 
the dust instrument on board the Galileo spaceprobe. 
Both dust instruments have been described in previous publications 
\cite{gruen1992a,gruen1992b,gruen1995a}. 

In the following we give a brief overview 
and an update of the most important scientific achievements of the Ulysses 
dust measurements. References to other works related to Ulysses and Galileo 
measurements on
dust in the Jovian system, interstellar dust in the heliosphere and on
the interplanetary meteoroid complex are given in various earlier publications
\cite{gruen1995c,gruen1995b,krueger1999a,krueger1999b}. For comprehensive 
reviews of the scientific achievements of the
Ulysses mission, including those from the dust investigations, the
reader is referred to {\em The heliosphere at solar minimum: The Ulysses
perspective} \cite{balogh2001} and the chapter on dust by Gr\"un et al.
\cite*{gruen2001} in the same book.

With the Ulysses dust instrument burst-like intermittent streams of tiny 
dust grains were discovered in interplanetary space \cite{gruen1993} 
which had been emitted from the Jovian system \cite{zook1996}. These 
grains strongly interact with the interplanetary and the Jovian magnetic 
fields \cite{horanyi1997,gruen1998}. After an ongoing debate 
about the origin of these grains (Io, Jupiter's gossamer ring, etc.,
Hamilton and Burns, 1993, Hor\'anyi et al., 1993),
\nocite{hamilton1993a,horanyi1993a} 
Io was recently confirmed as their ultimate source \cite{graps2000}.

Ulysses and Galileo dust measurements have been used to derive the 
3-dimensional distribution of the interplanetary dust complex 
and their relation to the underlying populations of parent bodies like 
asteroids and comets 
\cite{divine1993,gruen1997a}. Studies of asteroidal dust released from 
the IRAS dust bands show that 
they are not efficient enough dust sources to maintain a 
stable interplanetary dust cloud \cite{mann1996b}.
Recently, the properties of $\beta$ meteoroids 
(i.e. dust particles which leave the Solar System on unbound orbits due
to acceleration by radiation pressure) have been studied with the 
Ulysses dust data set \cite{wehry1999}. 

Another discovery made with the Ulysses detector were interstellar 
particles sweeping through the Solar System \cite{gruen1993}. 
The grains were identified by their impact direction and impact 
velocities, the latter being compatible with particles moving 
on hyperbolic heliocentric trajectories \cite{gruen1994a}. Their
dynamics depends on grain size and is stongly affected by the interaction
with the heliosphere and by solar radiation pressure 
\cite{mann2000,landgraf2000b}. This has a strong influence 
on the size distribution and fluxes of grains measured inside the 
heliosphere. The mass distribution of grains measured at 
heliocentric distances inside 4~AU shows a lack of grains with
masses $\rm \sim 10^{-13}\,g$: particles for which the ratio between 
the solar radiation force and the gravitational force, $\beta$, exceeds 
unity are rejected from the Sun and thus hindered from entering the 
inner Solar System ($\beta$ gap; \cite{landgraf1999}). This in turn 
allows to infer optical properties of the grains which are 
consistent with different models of dust composition and structure.
% astronomical silicates or organic refractory material.

Ulysses and Galileo in situ measurements imply that the intrinsic size 
distribution of interstellar grains in the local interstellar medium
extends to grain sizes larger than those detectable by astronomical observations
\cite{frisch1999,landgraf2000a,gruen2000b}. The existence of such 
'big' interstellar
grains is also indicated by observations of radar meteors entering
the Earth's atmosphere \cite{taylor1996b,baggaley2000}. The 
dust-to-gas mass ratio in the local interstellar cloud is several 
times higher than the standard interstellar value derived from 
cosmic abundances, implying the existence of inhomogeneities 
in the diffuse interstellar medium on relatively small length scales.

%For full reviews of the scientific achievements of the 
%Ulysses mission including results from the dust instrument, the 
%reader should consult {\em The Heliosphere at solar minimum: The Ulysses
%perspective} \cite{marsden2001}.
%References to other works related to Ulysses and Galileo measurements on 
%dust in the Jovian system, interstellar dust in the heliosphere and on 
%the interplanetary meteoroid complex are given in various publications 
%\cite{gruen1995c,gruen1995b,krueger1999a,krueger1999b}.

This is the seventh paper in a series dedicated to presenting both 
raw and reduced data obtained from the dust instruments on board the 
Ulysses and Galileo spacecraft. The reduction process of Ulysses and 
Galileo dust data has been described by Gr\"un et al. 
(1995c, \nocite{gruen1995a} hereafter Paper~I).
In Papers~III and V \cite{gruen1995c,krueger1999b} we present the 
Ulysses data set spanning the five year time period from October 1990
to December 1995. Papers~II and IV \cite{gruen1995b,krueger1999a}
discuss the six years of Galileo data from October 1989 to December 
1995. The current paper extends the Ulysses data set 
from January 1996 until December 1999, and a companion paper 
\cite{krueger2001a} (Paper~VI) presents Galileo's 1996 measurements.

The main data products are a table of the impact 
rate of all impacts determined from the particle accumulators and a table 
of both raw and reduced data of all dust impacts for which the 
full data set of measured impact parameters has been transmitted to Earth.
The information presented in these papers is similar to data which we 
are submitting to the various data archiving centres (Planetary 
Data System, NSSDC, Ulysses Data Centre). Electronic access to the 
data is also possible via the world wide web: http://www.mpi-hd.mpg.de/dustgroup/.

This paper is organised like Papers~III and V. We begin with an 
overview of important events of the Ulysses mission between 1996 and 1999 
(Section~\ref{mission}). Sections~\ref{events} and \ref{analysis} describe
and analyse the Ulysses dust data set for this period. 
In Sect.~\ref{discussion} we discuss the properties of different populations
of interplanetary and interstellar dust in the data set and compare our 
Ulysses measurements with dust fluxes measured in the outer Solar System 
with the Pioneer~10 and 11 spacecraft. In Sect.~\ref{summary} we 
summarise our conclusions. An appendix gives a conversion between
impact charges as measured by the
dust instrument and grain masses and sizes for different impact speeds.

\section{Mission and instrument operation} \label{mission}

\subsection{Ulysses Mission and Dust Instrument Characteristics}

\begin{figure}[h]
\epsfxsize=12.5cm
\epsfbox{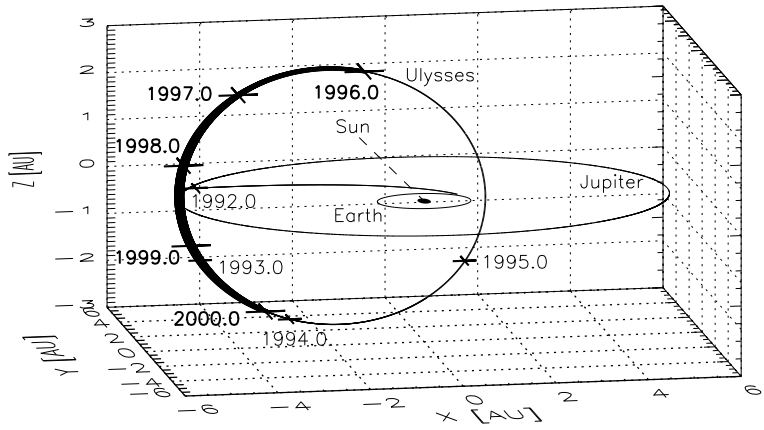}
%\epsfbox{../figures/trajectory.ps}
        \caption{\label{trajectory}
The trajectory of Ulysses in ecliptic coordinates. The Sun is in the centre. 
The orbits of Earth and Jupiter indicate the ecliptic plane. The initial
trajectory of Ulysses was in the ecliptic plane. Since Jupiter flyby 
in early 1992 the orbit has been almost perpendicular to the ecliptic plane 
(79$^{\circ}$ inclination).
Crosses mark the spacecraft position at the beginning 
of each year. The 1996 to 1999 part of the trajectory is shown as a 
thick line. Vernal equinox is to the right (positive x axis).
}
\end{figure}

The Ulysses spacecraft was launched on 6 October 1990. 
A swing-by manoeuvre at Jupiter in February 1992 rotated the orbital 
plane 79$^{\circ}$ relative to the ecliptic plane. On the resulting 
trajectory (Fig.~\ref{trajectory}) Ulysses passed under the south pole 
of the Sun (October 1994), crossed the ecliptic plane at a 
perihelion distance of 1.3~AU (March 1995) and passed over the 
Sun's north pole (August 1995). In April 1998 the spacecraft 
finished its first out-of-ecliptic orbit and  crossed 
the ecliptic plane again at an aphelion distance of 5.4~AU.
In 2000 and 2001 it will make its second pass over the Sun's polar 
regions. Orbital elements for the out-of-ecliptic part of the 
Ulysses trajectory are given in Table~\ref{orbital_elements}.

\begin{figure}
\epsfxsize=10.5cm
\epsfbox{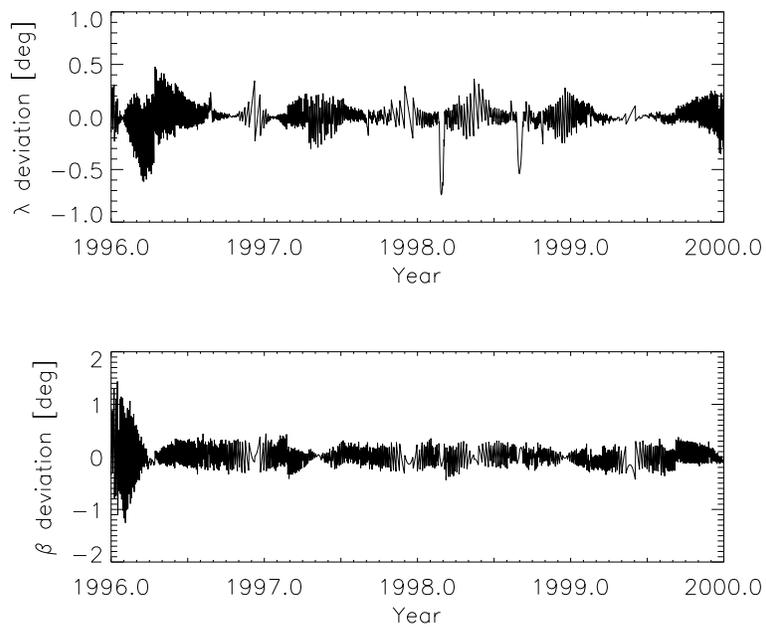}
%\epsfbox{../figures/pointing.ps}
        \caption{\label{pointing}
Spacecraft attitude: deviation of the antenna pointing direction 
(i.e. positive spin axis) from the nominal Earth direction. The angles are 
given in ecliptic longitude (top) and latitude (bottom, equinox 1950.0).
}
\end{figure}

Ulysses spins at five revolutions per minute about the centre line 
of its high gain antenna which normally points at Earth. 
Fig.~\ref{pointing} shows the deviation of the spin axis 
from the Earth direction for the period 1996 to 1999.
Most of the time the spin axis pointing was within 
$0.5^{\circ}$ of the nominal Earth direction. This small deviation 
is usually negligible for the analysis of measurements with the 
dust detector.
The Ulysses spacecraft and mission are explained in more detail by Wenzel 
et al. \cite*{wenzel1992}. Details about the data transmission to Earth 
can also be found in Paper~III.

The Ulysses dust detector (GRU) has a 140$^{\circ}$ wide field of view and
is mounted at the spacecraft nearly at right angles (85$^{\circ}$) to the antenna 
axis (spacecraft spin axis). Due to this mounting geometry, the dust sensor 
is most sensitive to particles approaching from the plane perpendicular to the 
spacecraft-Earth direction. The impact direction of dust particles is 
measured by the rotation angle which is the sensor viewing direction at the 
time of a dust impact. During one spin revolution of the spacecraft the 
rotation angle scans through a complete circle of 360$^{\circ}$. 
Zero degrees rotation angle is defined to be the 
direction closest to ecliptic north. At high ecliptic latitudes, however, the 
sensor pointing at 0$^{\circ}$ rotation angle significantly deviates from the 
actual north direction. During the passages over the Sun's polar regions the sensor 
always scans through a plane tilted by about $30^{\circ}$ from the ecliptic 
and all rotation angles lie close to the ecliptic plane 
(cf. Fig.~4 in Gr\"un et al., 1997\nocite{gruen1997a}). A sketch of the viewing 
geometry around aphelion passage can be found in Gr\"un et al. 
\cite*{gruen1993}.
 
Table~\ref{event_table} gives significant mission and dust instrument 
events from 1996 to 1999. Earlier events are only listed if especially significant.
A comprehensive list of events from launch until the end of 1995 is given in 
Papers~III and V. 

\subsection{Instrumental Noise}

Analysis of the in-orbit noise characteristics of the dust
instrument (Paper~III) led to a relatively noise-free configuration with which 
the instrument has been normally operated so far: 
channeltron voltage 1140~V (HV~=~3); event definition status such that either the 
channeltron or the ion-collector channel can, independent of each other,
start a measurement cycle 
(EVD~=~C,~I); detection thresholds for ion-collector, channeltron and 
electron-channel set to the lowest levels and the detection threshold for 
the entrance grid set to the first digital step (SSEN~=~0,~0,~0,~1). See 
Paper~I for a description of these terms. This instrument set up is hereafter
called nominal configuration.

Dedicated noise tests were performed at about one month intervals to monitor 
instrument health and noise characteristics. During all these tests the 
operational configuration was changed in four steps at one-hour intervals, 
starting from the nominal setting described above: 
a) set the event definition status such that the channelton, the ion 
collector and the electron-channel can initiate a measurement cycle 
(EVD~=~C,~I,~E);
b) set the thresholds for all channels to their lowest levels 
(SSEN~=~0,~0,~0,~0);
c) reset the event definition status to its nominal configuration (EVD~=~C,~I) 
and increase the channeltron high voltage by one digital step (HV~=~4);
d) reset the instrument to its nominal operational configuration (channeltron 
high voltage to HV~=~3, detection thresholds to SSEN~=~0,~0,~0,~1).
 
The noise tests revealed a long-term drop of the noise sensitivity of the 
instrument: in step c), when the channeltron high voltage was increased, fewer 
noise events were triggered compared to the earlier period before 1996. 
This is most likely caused by a reduction in the channeltron amplification 
due to ageing
(see also Section \ref{events}) although less  noise is generally expected 
far away from the Sun. In addition, the average charge amplitudes
measured at the target during noise events were reduced. This is probably 
an effect of degradation of the instrument electronics.

The noise response of the Ulysses dust detector will be continuously monitored 
in the future to maintain stable instrument operation. So far, the degradation 
does not lead to a serious reduction in the instrument sensitivity: although we 
expect a lower number of class~3 events due to the reduced channeltron 
amplification, the dust impacts causing these events should show up
as class~0 impacts. 

From 1996 to 1999 three spacecraft anomalies occurred during which all 
scientific instruments on board Ulysses were switched off automatically. 
These Disconnections of all Non-Essential Loads are called DNELs for short. 
Within about two days 
after each DNEL, the dust instrument was switched on again and reconfigured 
to its nominal operational mode. The resulting loss of continuous measurement 
time with the dust instrument is negligible.

Two heaters allow for a relatively stable operating temperature of the dust
sensor (heating powers 400~mW and 800~mW, respectively). Both 
heaters were switched on during the time period 1996 to 1999 considered here,
providing a constant heating power of 1,200~mW. Sensor heating was 
necessary because the spacecraft was outside 3~AU all the time and hence
received relatively little radiation from the Sun.
During DNELs the heaters were switched off, but switched on again several
hours before the instrument was reconfigured so that a stable operating
temperature could be achieved at switch-on to avoid instrument damage.
From 1996 to 1999 the temperature of the dust sensor was between 
-10$^{\circ}\,\rm C$ and -23$^{\circ}\,\rm C$ which is within the 
specified operational range. A switch-off of the 
400~mW heater planned for September 1999 for power saving reasons
on board the spacecraft did not become necessary. Due to a 
decreasing power output of the radioisotope batteries (RTGs) power 
saving issues will become important for instrument operation on board
Ulysses after 2001.

\begin{figure}
\epsfxsize=10.5cm
\epsfbox{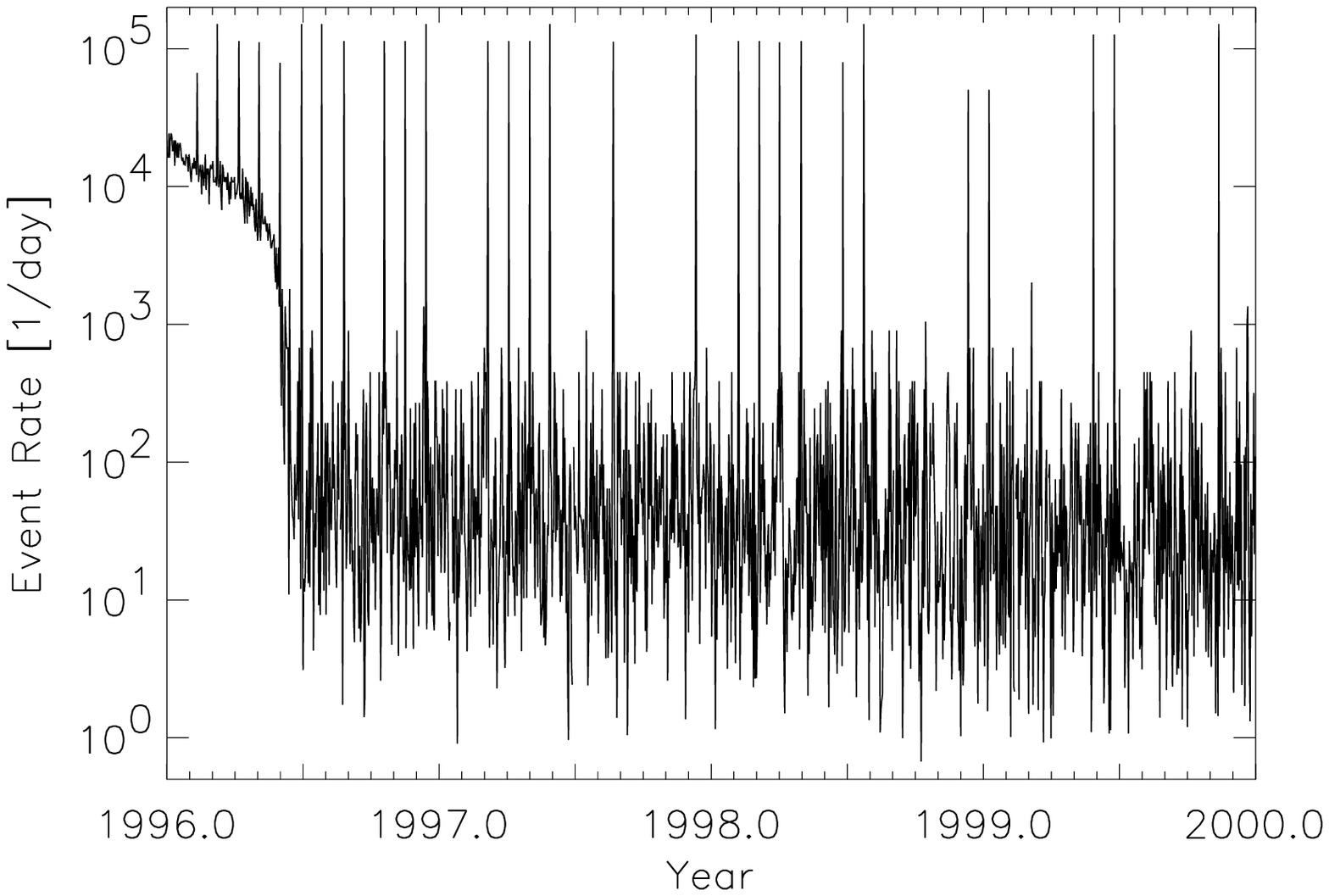}
%\epsfbox{../figures/noiserate1.ps}
\epsfxsize=10.5cm
\epsfbox{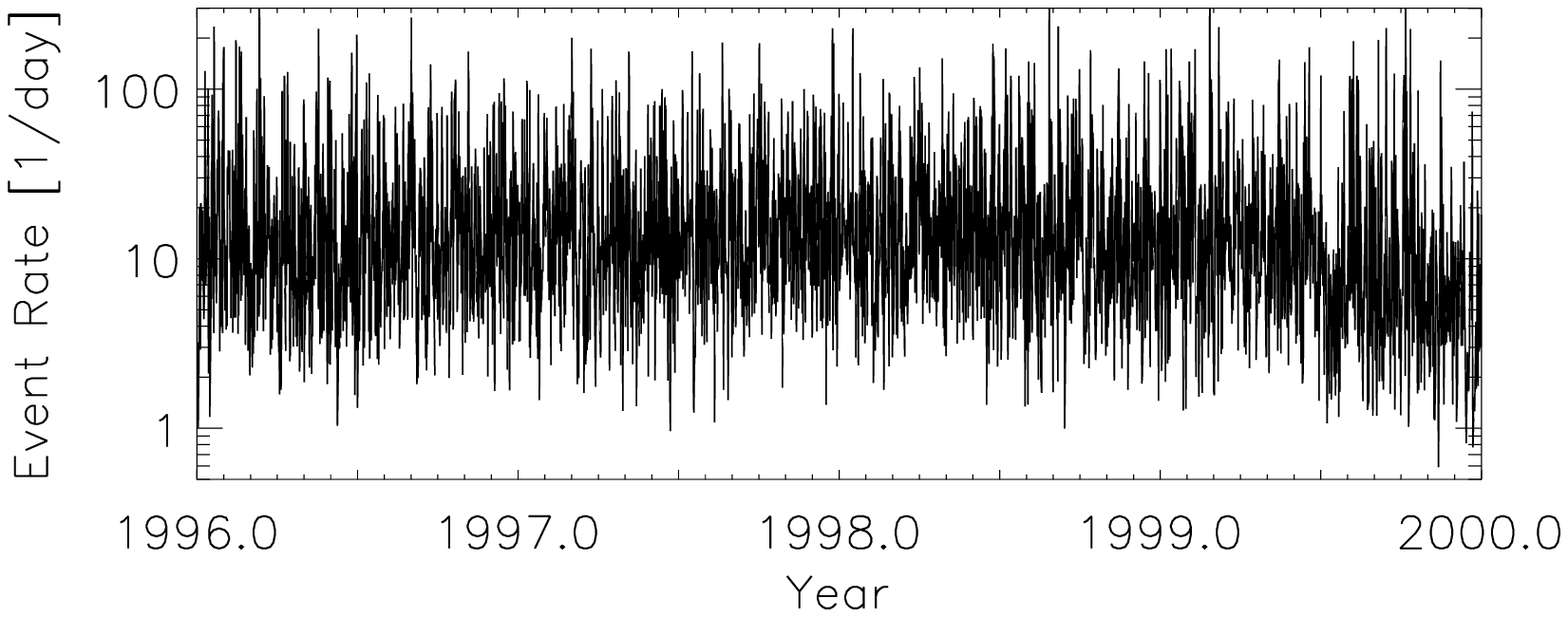}
%\epsfbox{../figures/noiserate2.ps}
        \caption{\label{noiserate}
Noise rate (class~0 events) detected with the dust instrument. 
Upper panel: Daily 
maxima in the noise rate (determined from the AC01 accumulator). The 
sounder was operated for only 2\,\% of the total time, and the daily 
maxima are dominated by sounder noise. Sharp spikes 
are caused by periodic noise tests and short periods of 
reconfiguration after DNELs (Table~\ref{event_table}). 
Lower panel: Noise rate detected during quiet intervals when the 
sounder was switched off, which was the case about 
98\,\% of the time. The curve shows a one-day average 
calculated from the number of AC01 events for which the complete 
information has been transmitted to Earth.
}
\end{figure}

Figure~\ref{noiserate} shows the noise rate of the dust instrument 
for the 1996 to 1999 period. 
The upper panel shows the daily maxima of the noise rate. They are dominated 
by noise caused by interference with the sounder of the Unified RAdio and Plasma 
wave instrument (URAP) on board 
Ulysses \cite{stone1992}. The sounder is typically operated for periods of
only 2~min with quiet intervals of about 2~hours. Thus, the high noise rates 
caused by the sounder occurred only during about 2\% of the total time. The 
remaining 98\% are free of sounder noise. The noise rates measured during 
sounder operation  are correlated with the distance to, and 
the position of, the Sun with respect to the sensor-viewing direction
\cite{baguhl1993}. Most noise events  were triggered when the Sun
shone directly into the sensor. Strong sounder noise occurred in 
1994 and 1995 while Ulysses was within about 3~AU heliocentric distance 
(Paper~V) and this continued until mid 1996 (Fig.~\ref{noiserate}, upper
panel). In this time period the sounder noise was sufficiently high to 
cause significant dead time in the dust instrument during the time
intervals of sounder operation.
 
When the spacecraft was at a
large heliocentric distance, the noise rate was very low even during periods 
of sounder operation. This was the case after mid 1996. 
Individual sharp spikes in the upper panel of Fig.~\ref{noiserate}
are caused by noise tests which occurred at monthly intervals. After mid 1996 
many of these noise tests did not show up in the diagram anymore because of 
degrading of the channeltron which led to a reduced noise sensitivity 
of the instrument (see also Sect.~\ref{analysis}).

The sounder was operated frequently during noise tests. In
step c) of the noise test (channeltron high voltage increased) with simultaneous
sounder operation, the noise signals usually showed a channeltron charge 
amplitude CA=1 (digital units) between 1991 and 1996. Later in the 
mission, values of CA=0 dominated which is also indicative of a channeltron 
degradation.

The noise rates when the sounder was switched off are shown in the lower panel of 
Fig.~\ref{noiserate}. The average was about 10 events 
per day and this was random noise not related to the sounder.  Dead time is 
negligible during these periods.

\section{Impact events} \label{events}

Impact events are classified into four classes and six ion charge amplitude 
ranges which lead to 24 individual categories. In addition, the instrument
has 24 accumulators with one accumulator belonging to one individual 
category. Class~3, our
highest class, are real dust impacts and class~0 are mostly noise events. 
Depending upon the noise of the charge measurements, classes~1 and 2 
can be true dust impacts or noise events. This classification 
scheme for impact events has been described in Paper~I and the 
scheme is still valid for the Ulysses dust instrument. In contrast to 
the Galileo dust instrument which had to be reprogrammed because of 
the low data transmission capabilities of the Galileo spacecraft, no such 
reprogramming has been necessary for Ulysses so far. Most of the data 
processing for Ulysses can be done on the ground.

Between 1 January 1996 and 31 December 1999 the complete data sets (sensor 
orientation, charge amplitudes, charge rise times, etc.) of 16,758 
events 
including 218 dust impacts were transmitted to Earth. Table~3 
lists the number of all dust impacts counted with the 24 accumulators of 
the dust instrument. `AC{\em xy}' refers to class number `$x$' and amplitude
range `$y$' (for a detailed description of the accumulator categories
see Paper~I). As discussed in the previous section, most noise events were 
recorded during the short time 
periods when either the sounder of the URAP instrument was operating 
(Paper~III) or when the dust instrument was configured to its high sensitive state 
for noise tests, or both. During these periods many events were only counted 
by one of the 24 accumulators
because their full information was overwritten before the data could be
transmitted to Earth. Since the dust impact
rate was low during times outside these periods, it is expected
that only the data sets of very few true dust impacts were lost. 

All 218 dust impacts detected from 1996 to 1999 for which the complete 
information exists are listed in Table~4. Dust particles are identified by their 
sequence number and their impact time (first two columns). The event category -- 
class (CLN) and amplitude range (AR) -- are given in the third and fourth 
columns. Raw data as transmitted to Earth 
are shown in the next columns: sector value (SEC) which is the
spacecraft spin orientation at the time of impact,
impact charge numbers (IA, EA, CA) and rise times (IT, ET), time
difference and coincidence of electron and ion signals (EIT, EIC),
coincidence of ion and channeltron signal (IIC), charge reading at
the entrance grid (PA) and time (PET) between this signal and
the impact. Then the instrument configuration is given: event
definition (EVD), charge sensing thresholds (ICP, ECP, CCP, PCP) and
channeltron high voltage step (HV). Compare Paper~I for further
explanation of the instrument parameters. 

The next four columns in Table~4 give information about Ulysses' orbit: 
heliocentric distance (R), ecliptic longitude and latitude (LON, LAT) 
and distance from Jupiter ($\rm D_{Jup}$, in astronomical units). The 
next column gives the 
rotation angle (ROT) as described in Sect.~\ref{mission}. 
%Whenever this value is unknown, ROT is arbitrarily set to
%999. This occurs 0 times. 
Then follows the pointing direction of the dust instrument at 
the time of particle impact in ecliptic longitude and latitude 
($\rm S_{LON}$, $\rm S_{LAT}$).
%When ROT is not valid $\rm S_{LON}$ and $\rm S_{LAT}$ are useless and are 
%also set to 999. 
Mean impact velocity ($v$, in $\rm km\,sec^{-1}$) and velocity error factor 
(VEF, i.e. multiply or
divide stated velocity by VEF to obtain upper or lower limits) as well 
as mean particle mass ($m$, in grams) and mass error factor (MEF) are given in the 
last columns. For VEF $> 6$, both velocity and mass values should be
discarded. This occurs for 21 impacts. No intrinsic dust charge values 
are given (see Svestka et al. \cite*{svestka1996} for a detailed analysis).
Recently, reliable dust charge measurements for interplanetary dust grains 
were reported for the Cassini dust detector. 
These measurements may lead to an improved unterstanding of the charge measurements
of Ulysses and Galileo in the future. 

In this paper we use different parameters to characterise particle 
sizes (radius, mass, impact charge, amplitude range). A conversion
between these parameters is given in the appendix.

\section{Analysis} \label{analysis}

\begin{figure}
\epsfxsize=8.0cm
\epsfbox{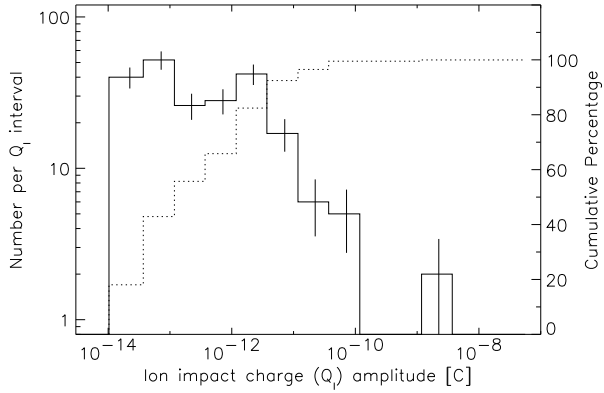}
%\epsfbox{../figures/nqi.ps}
        \caption{\label{nqi}
Distribution of the impact charge amplitude $Q_I$ for all 
dust particles 
detected from 1996 to 1999. The solid line
indicates the number of impacts per charge interval, and the 
dotted line shows the cumulative percentage. 
Vertical bars
indicate the $\rm \sqrt{n}$ statistical error. 
}
\end{figure}

The most important impact parameter determined by the dust instrument is
the positive charge measured on the ion collector, $Q_I$, 
because it is relatively insensitive to noise. Figure~\ref{nqi} shows the 
distribution of $Q_I$ for all dust particles detected from 1996 to 
1999. Ion impact charges have been detected over the entire range of six 
orders of magnitude in impact charge that can be measured by the dust instrument. 
The maximum measured charge was $Q_I \rm = 2 \cdot 10^{-9}\,C$, well below 
the saturation limit of $\rm \sim 10^{-8}\,C$. 
% and may thus constitute lower limits of the actual impact charges. 

In the earlier 1993 to 1996 data set (Fig. 4 in Paper~III) the impact charge 
distribution was reminiscent of three individual particle populations: 
small particles with impact charges $ Q_I \rm <10^{-13}\,C$ (AR1), intermediate
particles with $ \rm 10^{-13}\,C \leq \it Q_I \rm \leq 3\cdot 10^{-11}\,C $ 
(AR2 and AR3) and big particles with $Q_I \rm > 3\cdot 10^{-11}\,C$ 
(AR4 to AR6). This is also visible in the present data set, although 
somewhat less clear: The small and the intermediate population 
can be distinguished whereas the big population seems to be 
represented by only two impacts at about $2 \cdot 10^{-9}$ C and a few 
at $10^{-10}$ to $\rm 10^{-11}\,C$. The intermediate 
particles are mostly of interstellar origin and the big particles are 
attributed to interplanetary grains 
\cite{gruen1997a}, see also Sect.~\ref{discussion}). The small particles 
(AR1) occur mostly at high ecliptic latitudes over the polar regions of the Sun
and at low ecliptic latitudes around Ulysses' aphelion ecliptic 
plane crossing. Those at high ecliptic latitudes are attributed to a population of
interplanetary $\beta$-meteoroids \cite{baguhl1995b,hamilton1996,wehry1999}.

\begin{figure}
\epsfxsize=8.0cm
\epsfbox{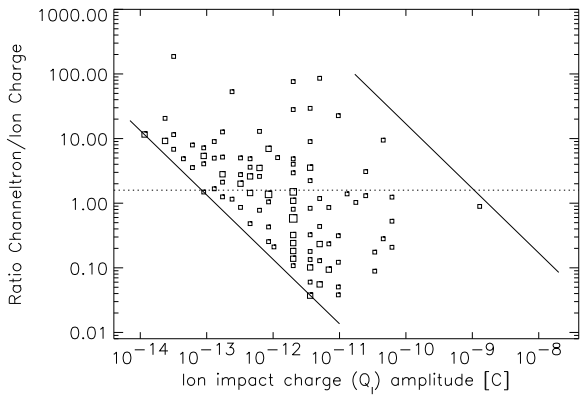}
%\epsfbox{../figures/qiqc.ps}
        \caption{\label{qiqc}
Channeltron amplification factor $A = Q_C/Q_I$
as a function of impact charge $Q_I$ for all dust impacts 
detected between 1996 and 1999. The solid lines denote the sensitivity
threshold (lower left) and the saturation limit (upper right) of the channeltron. Squares
indicate dust particle impacts. The area of each square is proportional to 
the number of events included (the scaling of the squares is the same 
as that in Paper~III). The dotted horizontal line shows the mean value 
of the channeltron amplification A\,=\,1.6 
for ion impact charges $\rm 10^{-12}~C < \it  Q_I < \rm 10^{-11}~C$.
}
\end{figure}

The ratio of the channeltron charge $Q_C$ and the ion collector
charge $Q_I$ is a measure of the channeltron amplification A, which
in turn is an important parameter for dust 
impact identification (Paper~I).
In Fig.~\ref{qiqc} we show the charge ratio $ Q_C/Q_I$ as a function of 
$ Q_I$ for the 1996 to 1999 dust impacts. In this time period 
the channeltron high voltage was set to 1140\,V (HV\,=\,3, nominal 
value) so that this diagram is directly comparable with similar 
diagrams in the previous Papers~III and V. Contrary to the previous
data sets there are only two impacts with $ Q_I \rm > 10^{-10}\,C$
and only few impacts occur along the
threshold line (left solid line). The paucity of data along the threshold
line is probably related to the long-term behaviour of the instrument.

The mean amplification determined from particles with $10^{-12}{\rm\,C} \le Q_I
\le 10^{-11}{\rm\,C}$ is $\rm A \simeq 1.6$ which is about 25\% below 
the value derived from the first five years of the mission
(Papers~III and V). This implies that the channeltron shows some 
degradation during the more than nine years since launch of the 
Ulysses spacecraft. Much more severe degradation was found for 
the Galileo detector during Galileo's orbital tour in the Jovian 
system which is probably related to the harsh radiation 
environment in the magnetosphere of the giant planet. 
The channeltron high voltage of the Ulysses detector will have to be 
increased in the future, as has 
already been done with the Galileo detector. The channeltron degradation
is the reason that not all noise tests show up as individual sharp spikes in 
Fig.~\ref{noiserate}.

\begin{figure}
\epsfxsize=9.5cm
\epsfbox{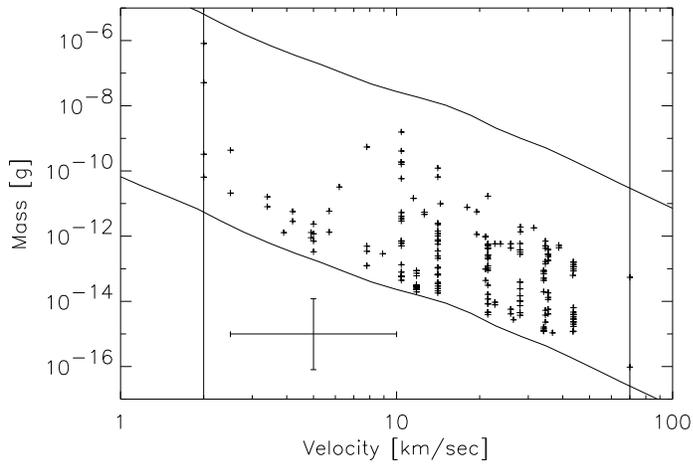}
%\epsfbox{../figures/mass_speed.ps}
        \caption{\label{mass_speed}
Masses and impact velocities of all impacts recorded with the Ulysses sensor 
from 1996 to 1999. The lower and upper solid lines indicate the threshold and
the saturation limit of the detector, respectively, and the vertical lines 
indicate the calibrated velocity range. A sample error bar is shown that 
indicates a factor of 2 uncertainty for the velocity and a factor of 10 for 
the mass determination.
}
\end{figure}

In Fig.~\ref{mass_speed} we show the masses and velocities of
all dust particles detected between 1996 and 1999. As in the
earlier period (1990 to 1995, Papers~III and V) velocities occur over
the entire calibrated range from 2 to $\rm 70\,km\,sec^{-1}$. The masses
vary over 10 orders of magnitude from $\rm 10^{-6}\,g$ to
$\rm 10^{-16}\,g$. The mean errors are a factor of 2 for the
velocity and a factor of 10 for the mass. The clustering
of the velocity values is due to discrete steps in the rise
time measurement but this quantisation is much smaller than the
velocity uncertainty. 
For many particles in the lowest two amplitude ranges (AR1 and
AR2) the velocity had to be computed from the ion charge signal
alone which leads to the striping in the lower mass range in 
Fig.~\ref{mass_speed} (most prominent above $\rm 10\,km\,sec^{-1}$). In the
higher amplitude ranges the velocity could normally be calculated
from both the target and the ion charge signal, resulting in a 
more continuous distribution in the mass-velocity plane. Impact 
velocities below about $\rm 3\,km\,sec^{-1}$ should be treated with caution
because anomalous impacts onto the sensor grids or structures 
other than the target generally lead to prolonged rise times and 
hence to unnaturally low impact velocities. 

\section{Discussion} \label{discussion}

In the time period considered in this paper the Ulysses spacecraft was 
beyond 3~AU from the Sun and at 
relatively low ecliptic latitudes most of the time
($\beta \leq 30^{\circ}$ after 1996). Figure~\ref{rate} shows the 
dust impact rate detected in various amplitude ranges together with the
total impact rate. In the 1996 to 1999 period the 
highest total impact rate ($\rm 3.5\cdot 10^{-6}\,sec^{-1}$) was 
recorded in early 1996 when Ulysses was at rather high ecliptic latitudes 
($\beta \sim 50^{\circ}$) about 3~AU from the Sun. Later, when the 
spacecraft reached lower latitudes further away from the Sun, 
the impact rate dropped by about a factor of 3. When Ulysses 
approached the ecliptic plane the impact rate increased again 
by a factor of 2 and reached a broad maximum around the aphelion ecliptic plane 
crossing (at 5.4~AU) in May 1998. The details of the impact rates 
measured in the various amplitude ranges will be discussed below.

Figure~\ref{rot_angle} shows the sensor orientation at the time of a 
particle impact (rotation angle, top panel). The expected impact directions
of interplanetary particles on bound heliocentric orbits (middle panel, 
\cite{gruen1997a}) and interstellar particles approaching from the 
interstellar upstream direction (bottom panel, \cite{witte1996})
are shown for comparison.
The intermediate sized and bigger particles (squares, 
impact charge $ Q_I \rm \geq 8 \cdot 10^{-14}~C$ which roughly corresponds 
to AR2-6) are clearly concentrated towards the interstellar 
direction. The small particles (crosses, $ Q_I \rm \leq 8 \cdot 10^{-14}~C$
roughly corresponding to AR1)
do not show such a strong concentration toward certain rotation 
angles, although many of them are also compatible with the interstellar
direction. The particles with the highest ion amplitude 
ranges (AR4 to AR6) are not distinguished in this diagram because 
they cannot be separated from interstellar particles by directional 
arguments alone. They have to be distinguished by other means
(e.~g. mass and speed). In addition, their total number is so small 
that they constitute only a small 'contamination' of the interstellar 
particles in Fig.~\ref{rot_angle}. 
The impact direction of the majority of the dust grains
detected outside 3~AU is compatible with particles of 
interstellar origin (bottom panel of Fig.~\ref{rot_angle}).

\subsection{Interstellar dust}
\label{interstellar_dust}

Interstellar particles are identified by their impact direction and
their impact speed: they approach from the same direction as the 
interstellar gas measured with Ulysses and move on hyperbolic 
trajectories through the Solar System \cite{gruen1994a,baguhl1995a,witte1996}.
In the Ulysses and Galileo data sets they are mostly found in 
amplitude ranges AR2 and AR3. 

\begin{figure}
\epsfxsize=10.5cm
\epsfbox{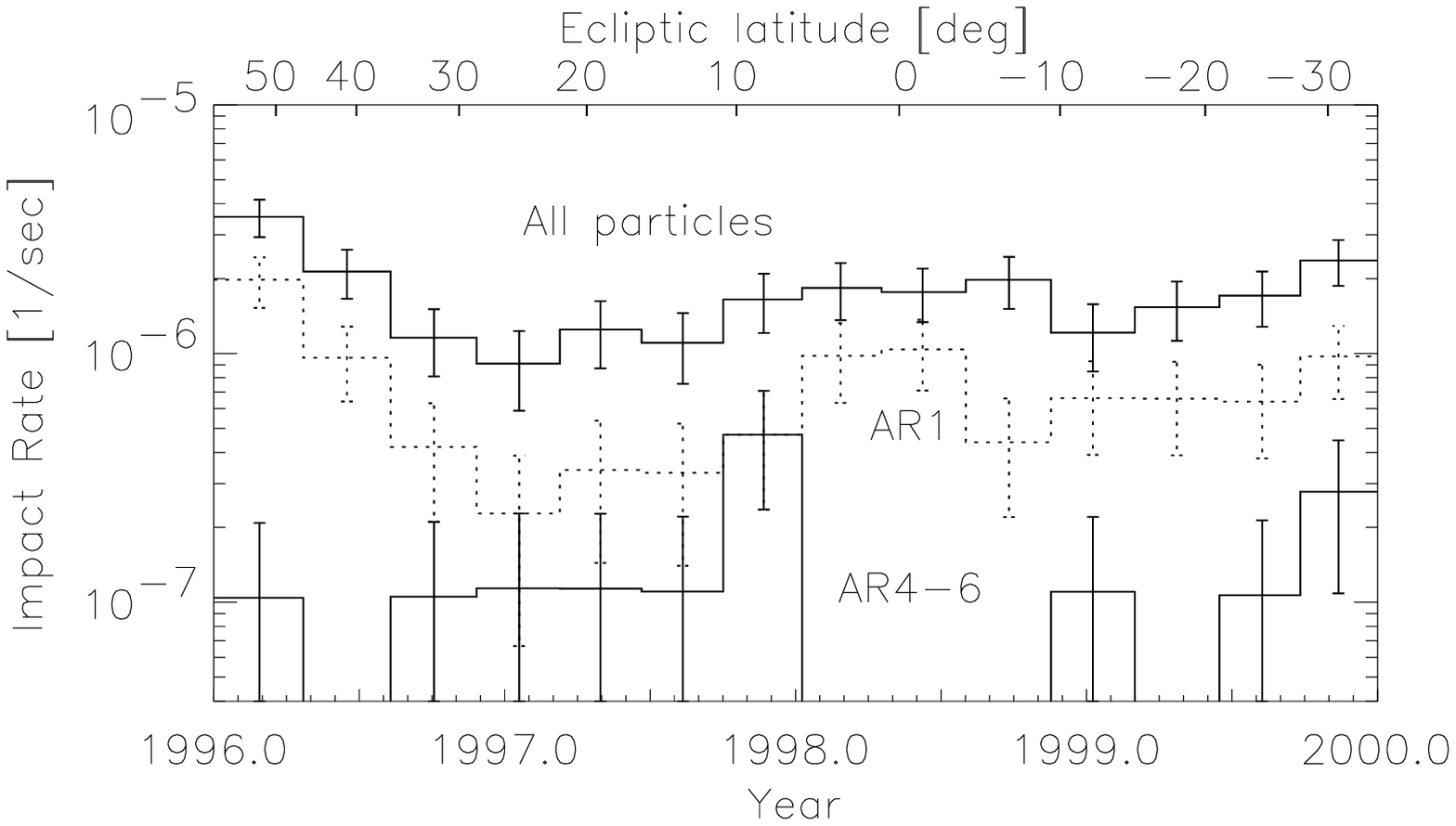}
%\epsfbox{../figures/rate1.ps}
\epsfxsize=10.5cm
\epsfbox{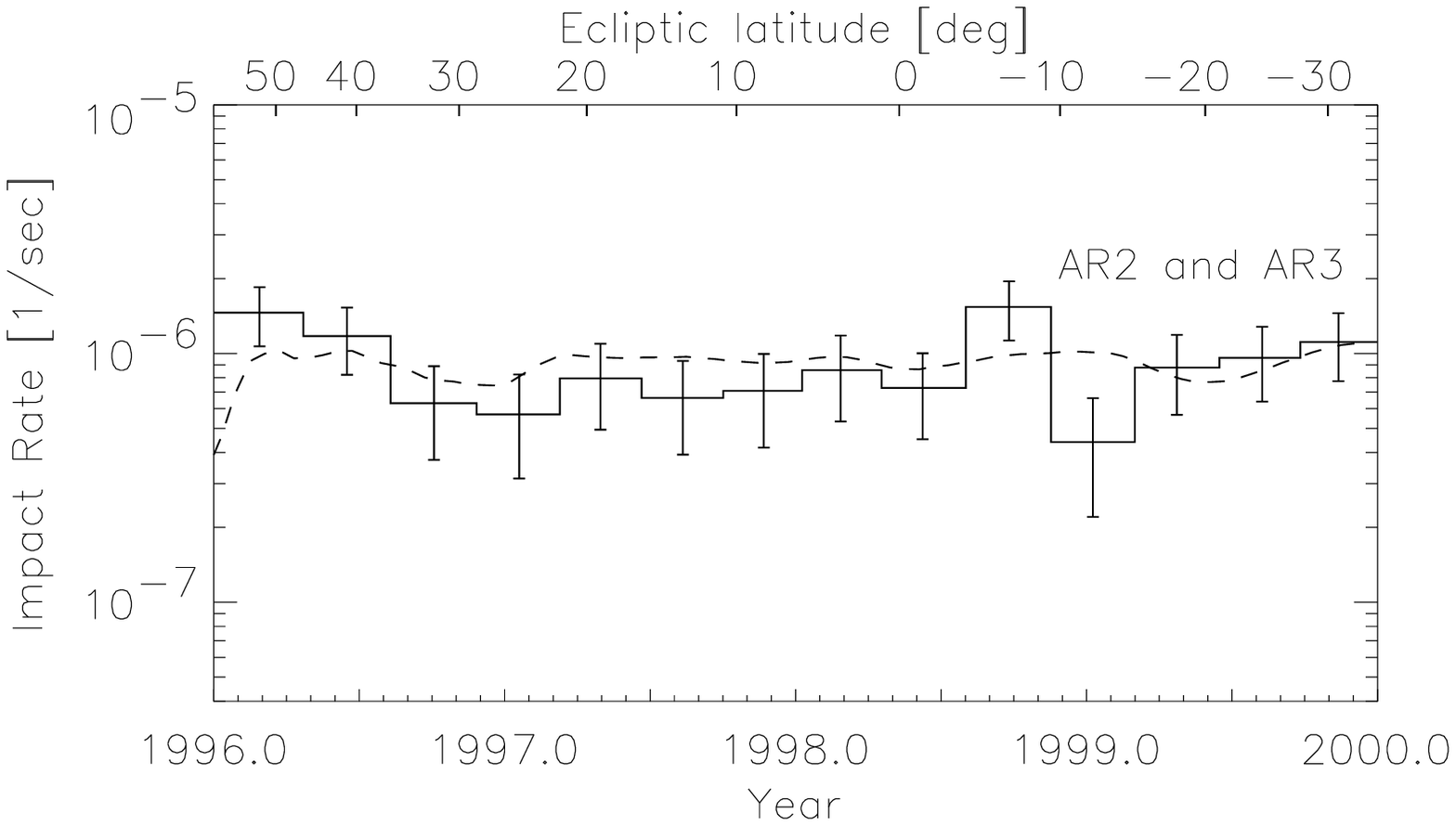}
%\epsfbox{../figures/rate2.ps}
        \caption{\label{rate}
Impact rate of dust particles detected with the Ulysses dust sensor 
as a function of time. The ecliptic latitude of the spacecraft is
indicated at the top. Upper panel: total impact rate (upper solid histograms),
impact rate of small particles (AR1, dotted histograms), and impact
rate of big particles (AR4 to AR6, lower solid histograms).
Note that a rate of about $\rm 1.0 \cdot 10^{-7}$ impacts per second 
is caused by a single dust impact in the averaging interval of about 
110 days.  Lower panel: impact rate of intermediate size particles 
(AR2 and AR3, solid histograms). A model for the rate of interstellar 
particles assuming a constant flux is superimposed as a dashed line. 
Vertical bars indicate the $\rm \sqrt{n}$ statistical error. 
}
\end{figure}

\begin{figure}
\vspace{-2cm}
\epsfxsize=15.5cm
\epsfbox{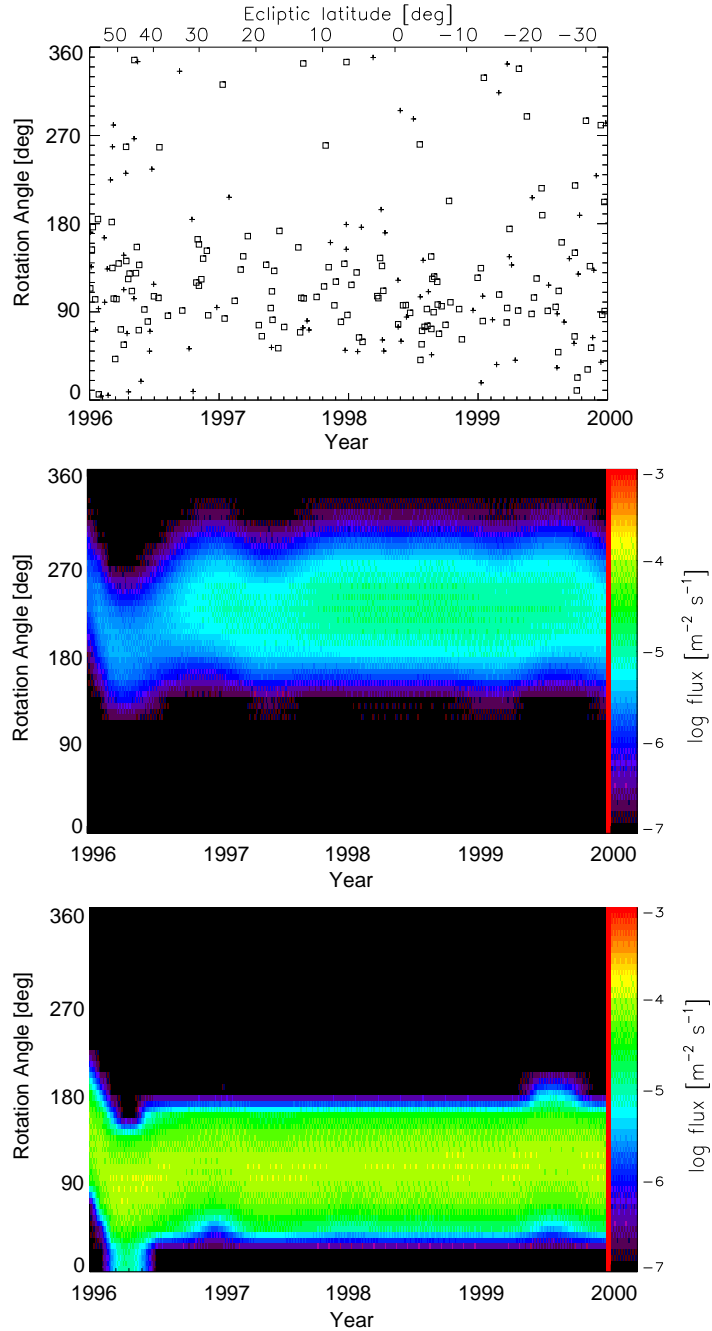}
%\epsfbox{../figures/rot_angle.ps}
        \caption{\label{rot_angle} 
Rotation angle vs. time. Top panel: all particles detected between 1996 
and 1999. Plus signs indicate particles 
with impact charge $Q_I \rm < 8 \cdot 10^{-14}$~C, squares those with
$ Q_I \rm \geq 8 \cdot 10^{-14}$~C.  
Ulysses' ecliptic latitude is indicated on the top. 
Middle panel: expected impact 
direction for interplanetary particles on heliocentric circular 
bound orbits concentrated towards the ecliptic plane (Gr\"un et al., 1997).
Bottom panel: expected impact direction for interstellar grains
approaching from the interstellar upstream direction (Witte et al., 1996).
The directions of most of the particles with larger $ Q_I$ are
consistent with the interstellar upstream direction. 
}
\end{figure}

The measured impact rate of grains in AR2 and AR3 is 
shown in the bottom panel of Fig.~\ref{rate}. 
As in the previous 1993 to 1995 period the impact rate of these intermediate
size grains was relatively constant, except a slight drop after mid 1996. 
However, on average the impact rate is 
lower than previously: after 1996 we find an average rate of 
$\rm \sim 8 \cdot 10^{-7}\,sec^{-1}$ whereas between 1993 and 1995 the rate was 
$\rm \sim 2 \cdot 10^{-6}\,sec^{-1}$.  

Figure~\ref{rate} also shows the expected impact rate of interstellar 
particles assuming that they approach from the direction of interstellar 
helium (Witte et al. 1996) and that they move through the Solar
System on straight trajectories with 
a relative velocity of $26\ {\rm km}\ {\rm s}^{-1}$.
This assumption means dynamically that radiation pressure
cancels gravity for these particles ($\beta=1$) and that their 
Larmor radii are large compared with the dimension of the Solar System. 
Both assumptions are reasonable for particles with masses between 
$10^{-13}$ and $10^{-12}\ {\rm g}$ which is the dominant 
size range measured for interstellar grains \cite{gruen1997a}.
The variation predicted 
by the model is caused by changes in the instrument's viewing 
direction with respect to the approach direction of the particles 
and changes in the relative velocity between the spacecraft and 
the particles. 
The dust particle flux is independent of heliocentric distance 
in this simple model, which gives relatively good agreement with the
observed impact rate. The assumed flux of oncoming particles at 
infinity, however,  had to be scaled down by a factor of 3  as
compared to a value of $ \rm 1.5 \cdot 10^{-4}\, m^{-2}\, sec^{-1}$ 
found before 1996 \cite{gruen1994a}. 

Detailed modelling of the 
interaction of the dust grains with the heliosphere \cite{landgraf2000b} 
shows that the change of the polarity of the interplanetary magnetic field 
with the solar cycle imposes a temporal variation of the flux  and the 
spatial distribution of small interstellar dust grains inside the 
heliosphere (radius smaller than $\rm 0.4\,\mu m$): the drop by almost a 
factor of 3 in the dust flux after mid 1996 is caused by a defocussing
effect of the interplanetary magnetic field which effectively prevents
small interstellar grains from entering the inner heliosphere.
This dynamical model better explains the observed 
impact rate between 1990 and 2000, including the 
drop after mid 1996. 

The mass distribution of interstellar grains entering the 
heliosphere is strongly modified by the interaction with the 
solar radiation field. Radiation pressure is strongly grain size 
dependent and even exceeds solar gravity for particle masses between 
about $10^{-16}$ and $\rm 10^{-14}\,g$. Since radiation pressure
counteracts solar gravity, the size distribution of
interstellar grains gets modified: Ulysses and Galileo measurements 
between 2 and 4~AU show a depletion in the range $\rm 10^{-14}$ to
$\rm 3 \cdot 10^{-13}\,g$ ($\beta$ gap \cite{landgraf1999}).  
This corresponds to a diameter range 0.2 to $\rm 0.6\,\mu m$ for 
spherical particles with a mass density of $\rm 2.5\,g\,cm^{-3}$.

\subsection{Submicrometre-sized interplanetary dust}

The impact rate of the smallest particles (AR1), and the variability
of that rate with time, are both much greater than those for the 
medium-sized AR2 and AR3 particles.
Particles at high ecliptic latitudes
are attributed to a population of small 
interplanetary particles on escape trajectories from the Solar 
System ($\beta$-meteoroids, Baguhl et al., 1995b, Hamilton et al., 
1996). \nocite{baguhl1995b,hamilton1996}
$\beta$-meteoroids have been detected with the Ulysses instrument
over the Sun's poles in 1994 and 
1995 \cite{wehry1999}. Due to the detection geometry, however, they
became undetectable after mid
1996. This is consistent with the drop in the impact rate by a factor
of 10 between January and December 1996 seen in Fig.~\ref{rate} 
(see Paper~V for the impact rate above the Sun's poles).
At the aphelion ecliptic plane crossing in 1998 a maximum 
occurred in the AR1 impact rate while Ulysses was within $\pm 5^{\circ}$ 
ecliptic latitude. This is consistent with interplanetary particles 
concentrated toward the ecliptic plane. 

Streams of tiny dust particles were
detected with the dust detectors on board Ulysses and Galileo out to a 
distance of 2~AU from Jupiter \cite{gruen1993,gruen1996b}. 
These grains originate from the Jovian system 
\cite{zook1996,gruen1998,graps2000}.
The dust stream particles impacts occurred only in the lowest amplitude range
AR1 (Papers~III and IV) and in principle they could contribute to 
the AR1 rate shown in Fig.~\ref{rate}. From 1996 to 
1999, the distance between Jupiter and Ulysses was 
more than 6~AU (Table~4) and around Ulysses' ecliptic plane crossing 
in 1998 the distance from Jupiter was 10~AU: Jupiter and Ulysses 
were on opposite sides of the Solar System and, hence, the 
Jupiter stream particles could not contribute to the AR1 impacts.

In February 2004 Ulysses will fly by Jupiter 
within a distance of 0.8~AU. This will again give the 
opportunity to measure the Jovian dust streams and to test 
presently existing models. (Note that the masses and velocities for the 
stream particles given in Papers II and III are not correct, Zook et al., 
1996; 
\nocite{zook1996} 
in reality the stream particles are smaller and 
faster than implied by the instrument calibration). 

\subsection{Micrometre-sized interplanetary dust; comparison with Pioneer~10 
and Pioneer~11 measurements}

The impact rate of big particles in AR4 to AR6 ($  Q_I \rm > 10^{-10}~C$) 
showed a clear
maximum at Ulysses' perihelion ecliptic plane crossing in 1995
(heliocentric distance 1.3~AU, Paper~V). These impacts were attributed 
to particles on low inclination heliocentric orbits \cite{gruen1997a}. From 
1996 to 1999, impacts in AR4 to AR6 occurred at a low and 
relatively constant rate of 1 to $\rm 2\cdot 10^{-7}\,sec^{-1}$,
except when Ulysses was close to the ecliptic plane 
(below $-5^{\circ} \lesssim \beta \lesssim 5^{\circ}, \rm \approx 5 \,AU$): 
in this period the dust detector recorded no impacts at all. 
Here one has to keep in mind that the statistical uncertainty 
is very large. In fact, many of the histogram bins for AR4-6 in Fig.~\ref{rate}
represent only one dust impact so that -- given the statistical 
uncertainty -- this is still consistent with a dust 
population concentrated toward the ecliptic plane. However, the 
measured impact directions of some of these grains are not in agreement
with low-inclination circular orbits about the Sun (Fig.~\ref{rot_angle}).
The impact directions of $\rm 180^{\circ} \leq ROT \leq 360^{\circ}$
rather indicate that the particles moved on highly inclined orbits.

The only spacecraft other than Ulysses and Galileo which were 
equipped with in situ dust detectors and traversed the outer Solar System were 
Pioneer~10 and 11. Both dust detectors consisted of an assembly of pressurized
cells and a device to monitor the pressure in each cell \cite{humes1980}. 
When a meteoroid penetrated the cell wall, the gas escaped from the cell
and the loss of pressure was detected. Once penetrated, a cell was no longer 
sensitive to meteoroid impacts, which lead to a decrease of the sensor 
area with time. The only difference between the Pioneer~10 and 11 detectors 
was that the stainless steel cell walls were $\rm 25 \mu m$ thick for the 
Pioneer~10 detector and $ \rm 50\,\mu m$ thick for the Pioneer~11 detector.
This difference in wall thickness provided a difference in threshold 
sensitivity of the detectors. The threshold sensitivities for 
meteoroids with a density of $\rm 0.5\, g\, cm^{-3}$ at impact speeds 
of $\rm 20\,km\,sec^{-1}$ was $\rm 8 \cdot 10^{-10}\,g$ for Pioneer~10 
and $\rm 6 \cdot 10^{-9}\,g$ for Pioneer~11, respectively. 

Between 3 and 18~AU from the Sun, measurements 
by the Pioneer~10 detector showed 
a dust flux of $\rm \approx 3 \cdot 10^{-6}\, m^{-2}\,sec^{-1}$ 
out to 18~AU \cite{humes1980}. At this distance
the nitrogen in the pressurized cells froze out and measurements
further away from the Sun became impossible. Dust measurements by Pioneer~11 
were obtained out to Saturn's orbit. During three passages of 
Pioneer~11 through the heliocentric distance range from 3.7 to 5~AU 
the detector observed a roughly constant dust flux between 
$\rm 3 \cdot 10^{-7}$ and $\rm 2 \cdot 10^{-6}\, m^{-2}\,sec^{-1}$. 
The absolute value is uncertain because the two measurement
channels of the dust instrument gave fluxes which differed by a factor 
of 5 \cite{humes1980}. On Pioneer~10, the second measurement channel 
of the dust instrument failed immediately after launch, so only one channel 
remained. Given the Pioneer~11 discrepancy, Pioneer~10
fluxes are also probably uncertain by at least a factor of 5. 

The Pioneer~10 and 11 dust measurements led to the 
conclusion that the dust had to be on highly eccentric orbits
with random inclinations rather than circular orbits concentrated 
towards the ecliptic plane. The distribution of the semi-major axis $a$
was $ n(a) \propto a^2$ if the particles were on bound orbits about 
the Sun \cite{humes1980}. 

We can compare the dust fluxes reported by Humes \cite*{humes1980} from the 
Pioneer~10 and 11 measurements between 3 and 5~AU with 
our Ulysses measurements in the same spatial region. If the inclinations
of the dust grains were randomly distributed, Ulysses should have detected 
these large 'Pioneer' particles independent of the ecliptic latitude of 
the spacecraft, similar to the interstellar dust grains. Since the 
interstellar particles form a monodirectional stream approaching 
from rotation angles $90\pm90^{\circ}$ (Fig.~\ref{rot_angle}), 
non-interstellar particles approaching from this direction are very 
difficult to separate. Therefore, we 
only consider rotation angles $270\pm 90^{\circ}$ where no interstellar
grains are expected to come from. This corresponds to half a rotation 
period of Ulysses. 

For an isotropic flux of dust grains, the sensor area of the Ulysses dust
detector spin-averaged over one hemisphere is about $\rm 300\,cm^{2}$.
Assuming an isotropic flux of $\rm 3 \cdot 10^{-6}\, m^{-2}\,sec^{-1}$
suggested by Pioneer~10, within four years of measurement time, Ulysses
should have detected about 10 impacts of particles with $m \rm \geq 8
\cdot 10^{-10}\,g$. Pioneer~11, sensitive to larger particles, predicts
that between 1 and 7 particles with $m \rm  \geq 6 \cdot 10^{-9}\,g$ should
be found by Ulysses. The Ulysses data in Table 4, however, shows far fewer
grains of this size with the right impact direction; only the largest two 
grains no. 1569 and 1575 with masses of
$\rm 2 \cdot 10^{-9}$ and $\rm 5 \cdot 10^{-8} \, g $, respectively, 
are sufficiently large.  
Even after accounting for the factor of 10
\cite{gruen1995a} uncertainty in the Ulysses mass determination, the
Pioneer and Ulysses rates are not in agreement. The fluxes of large
particles in the size range of $\rm \approx 10\, \mu m$ ($m \approx
\rm 2 \cdot 10^{-9}\,g$) implied by the Pioneer measurements are not seen
by Ulysses.
%For an isotropic flux of dust grains the sensor area of the 
%Ulysses dust detector spin-averaged over one hemisphere is 
%about $\rm 300\,cm^{2}$. Assuming an isotropic flux of 
%$\rm 3 \cdot 10^{-6}\, m^{-2}\,sec^{-1}$ (Pioneer~10), within four years 
%of measurement time Ulysses should have detected 11 impacts of 
%particles with $m \rm  \geq 8 \cdot 10^{-10}\,g$. Comparison with
%Table~4 shows that only 2 particles of this size and the right 
%impact direction have been recorded (grains no. 1569 and 1575).
%If we take into account the uncertainty in the mass determination 
%with the Ulysses dust detector which is a
%factor of 10 \cite{gruen1995a}, we find 4 particles with 
%$m > \rm 2 \cdot 10^{-10}\,g$ (grains no. 1569, 1575, 1585, 1634).
%If we take the fluxes measured by Pioneer~11, between 1 and 7 particles 
%with $\rm m \geq 6 \cdot 10^{-9}\,g$ should be found in the 
%Ulysses data set. However, only one particle in the required mass range 
%was recorded (no. 1575), whereas 2 particles are present if we
%take into account the uncertainty in the mass calibration (no. 1569 and 1575).
%To conclude, the fluxes of large particles in the size range of 
%$\rm \approx 10\, \mu m$ as implied by the Pioneer measurements cannot be 
%reproduced with Ulysses.

There are at least three possibilities to explain this discrepancy:
\begin{itemize}
\item[\rm A)] the calibrations of the Ulysses and Pioneer detectors are not 
correct by very large factors; 
\item[\rm B)] many of the Pioneer 'penetrations' 
are spurious and not due to actual meteoroid impacts; 
\item[\rm C)] there is 
a population of meteoroids that can easily be detected with the Pioneer
sensors, but not with the Ulysses sensor. 
\end{itemize}

To investigate possibility A),
we can ask which mass threshold of the 
Ulysses detector we have to assume in order to reproduce the dust 
fluxes observed by Pioneer~10/11. If we take a mass threshold of 
$ \sim \rm 10^{-13}\,g$ we find 12 particles with rotation angles
$270\pm 90^{\circ}$ and for a threshold of $ \sim \rm 10^{-12}\,g$
we find 5 particles in the Ulysses data set. Thus, only if 
we assume that the mass calibration of either the Ulysses or the 
Pioneer detectors is wrong by at least 3 orders of magnitude can we
match the fluxes observed by Pioneer and by Ulysses.

How do these numbers compare with the earlier data when Ulysses 
was also at large ecliptic latitudes between Jupiter fly-by in 
1992 and the passages above the Sun's polar regions in 1995? 
Between  mid 1992 and mid 1994 two particles were 
detected with masses larger than the Pioneer~10 threshold and
one particle larger than the Pioneer~11 threshold \cite{krueger1999b}, 
whereas 6 and between 0.6 and 4, respectively, should have been detected 
in order to reproduce the fluxes measured by the Pioneer detectors. 
Or vice versa, if we vary the detection threshold, the discrepancy 
is again roughly 3 orders of magnitude.

It should be noted that for Pioneer the calibrated mass of grains 
that can penetrate the cell walls of the detectors is only valid for 
impact speeds of $v = \rm 20\,km\,sec^{-1}$ and grain densities 
$\rm \rho = 0.5\,g\,cm^{-3}$. The threshold mass 
is roughly proportional to  $\rm \rho^{-0.5}$ and $v^{-2.5}$
\cite{humes1980}. Thus, an uncertainty in the grain density of 
a factor of 4 causes a change in the mass threshold of less than a 
factor of 2.  A decrease of the relative speed from $\rm 20\,km\,sec^{-1}$
at 1~AU to $\rm 10\,km\,sec^{-1}$  at 3 to 5~AU would increase the 
mass threshold by more than a factor of 5. Given the speed 
dependence of the mass threshold of the Ulysses detector which is
proportional to $v^{-3.5}$, this would somewhat reduce the order 
of 3 discrepancy found above. 

One might argue that erosion of the cell walls of the Pioneer detectors 
by particles smaller than the detection threshold has lead to a continuous 
degradation of the detection threshold. The area of each pressurised 
cell was $\rm 24.5\,cm^2$ whereas sizes of impact craters created by 
particles of micrometer sizes would have been in the range of a few 
ten micrometers. Thus, it would need an extremely large number of 
particle impacts below the threshold until one has a high probability
that a particle too small to penetrate the undamaged cell wall will
hit the cell in an eroded region. Given the low number of impacts
in interplanetary space, this process is extremely unlikely to happen.

For the Ulysses instrument one should note that the sensor
was only calibrated for masses up to $\rm 10^{-12}\,g$ 
at impact speeds of $\rm 10\,km\,sec^{-1}$ and $\rm 10^{-10}\,g$ 
at $\rm 2\,km\,sec^{-1}$, respectively. For larger masses the calibration 
had to be extrapolated \cite{gruen1992a}. However, the validity of 
this extrapolation is supported by calibration measurements of the 
DIDSY instrument on board the Giotto spacecraft \cite{goeller1987}.

Regarding possibility B) the factor of five 
discrepancy between the two measurement channels of the Pioneer~11 
dust detector is not understood. Furthermore, one channel of 
the Pioneer~10 instrument failed completely so that only one 
channel was left. Although, apart from this failure the Pioneer~10 instrument
performed very well, the Pioneer 10/11 data
should be treated with considerable caution.

For case C),
we have to consider the fields-of-view of the dust detectors on 
Pioneer 10/11 and on Ulysses: Ulysses has the 
highest sensitivity for dust particles approaching from a direction
more or less perpendicular to the spacecraft-Earth line whereas the 
Pioneer detectors have the highest sensitivity for grains approaching 
along the spacecraft-Earth line from the anti-Earth direction. 
Because of these very different detection geometries, Pioneer 10/11 may have detected 
dust grains on highly eccentric heliocentric orbits which were hardly 
detectable with Ulysses. In addition, 
the field-of-view of the Ulysses detector is $140^{\circ}$ 
and that of the Pioneer detector arrays is more than $180^{\circ}$ 
\cite{humes1980}. Depending on the distribution of the orbital elements 
of the grains this can lead to significant differences in the detected 
fluxes. For an incident isotropic flux  the expected difference is only about 
a factor of 2, whereas for particles on more or less radial heliocentric trajectories
the difference may be much larger: grains approaching the Sun on highly 
eccentric orbits are easily detected with the Pioneer detectors but hardly with 
Ulysses.

The Pioneer 10 and 11 spacecraft also carried photopolarimeters on board which
were used to map the zodiacal light and background starlight during the cruise 
to Jupiter in two optical bandpasses at 0.44 and $0.64\,\rm\mu m$. Beyond the 
asteroid belt no decrease in brightness with distance was detected, indicating
that the spacecraft were beyond the main zodiacal dust cloud and that the 
background starlight was directly measured. Only if the dust particles were 
darker than cometary dust grains could the measured brightnesses 
explain the in situ dust fluxes \cite{kneissel1990}.

It should also be noted that Pioneer~10 and 11 detected a few dust impacts during
Jupiter fly-by \cite{humes1976}. A more detailed analysis of these measurements 
may lead to clarification of the discrepancy between Ulysses and Pioneer~10/11 
discussed here.

What is the source of interplanetary dust grains in the outer Solar
System?
A significant source for interplanetary dust probably lies 
beyond Neptune's orbit: 
mutual collisions between Edgeworth-Kuiper belt objects (EKOs) plays a 
significant role as a dust production mechanism \cite{stern1996}. 
Impacts of interstellar grains onto the surfaces of (EKOs) may also lead to 
significant secondary dust ejection \cite{yamamoto1998}. 
Once released from their parent 
bodies the grains spiral towards the Sun under Poynting-Robertson and 
solar wind drags and may even reach Earth \cite{liou1996}. Particles
several micrometre in radius are most efficiently transported towards 
the Sun. Such grains should show up in AR4-6 in the Ulysses data. They 
should, however, show a flux increase towards the ecliptic plane.
Another dust source in the outer Solar System are long period 
comets. The particles they release into interplanetary space 
should move on highly inclined orbits with random inclinations.
In principle, such grains can support the interpretation that 
the grains detected by the Pioneer detectors move on highly eccentric
orbits, although the 
flux should increase towards the inner Solar System which is not 
seen. It should be noted that Ulysses also detected impacts of
big particles at high ecliptic latitudes above the Sun's polar
regions \cite{krueger1999b} which are compatible with highly
inclined orbits. 
Detailed modelling of the dust environment
in the outer Solar System including an investigation of different possible
source types is forthcoming (Landgraf et al., in prep.). 
The region between Jupiter and Saturn 
will be traversed by the Cassini spacecraft between 2001 and 2004 and the
dust distribution will be measured with the Cosmic Dust Analyser carried on 
board \cite{srama2001}. Together with modelling of the dust dynamics this 
will hopefully improve our understanding of the interplanetary dust complex 
beyond the asteroid belt.

\section{Summary} \label{summary}

In this paper, which is the seventh in a series of Ulysses and Galileo 
papers, we present data from the Ulysses dust instrument for the 
period January 1996 to December 1999 when Ulysses was traversing 
interplanetary space between 3 and 5~AU from the Sun at ecliptic
latitudes between $+50^{\circ}$ and $-35^{\circ}$. A total number of 218 
dust impacts has been recorded during this period. Together with 1477 
impacts recorded in interplanetary space and near Jupiter 
between Ulysses' launch in October 1990 and December 1995 
\cite{gruen1995c,krueger1999b}, the complete data set of dust impacts
measured with the Ulysses dust detector so far comprises 1695 
impacts. 

A relatively constant impact rate 
of 0.2 dust impacts per day was detected during 
the whole four-year time span. Most of these particles were of 
interstellar origin, a minor fraction being interplanetary 
particles.  

The flux measured with the Ulysses detector has been compared 
with the dust measurements obtained with the Pioneer~10 and 
11 detectors beyond 3~AU heliocentric distance \cite{humes1980}. 
The flux of particles with $ m \rm \geq 10^{-9}\,g$ as measured 
with Ulysses is about a factor of 5 lower than expected from the 
Pioneer~10/11 measurements. Although we are dealing with only a 
small number of impacts, the most plausible explanations are a)
the mass calibrations of the Ulysses and/or Pioneer detectors for
$\rm \approx 10\,\mu\,m$ grains differ by up to 3 orders of magnitude;
b) many of the Pioneer detections are 
spurious and not due to actual meteoroid impacts; c) the dust 
grains detected by Pioneer belong to a population of dust which 
could not or only partially be detected with Ulysses. 

%3) The impact rate of interstellar grains was rather constant
%between January 1995 and December 1996. A drop in the impact 
%rate after mid 1996 
%can be explained by a change in polarity of the interplanetary 
%magnetic field \cite{landgraf2000b}.

%4) At high ecliptic latitudes small dust particles have been 
%detected which are compatible with $\beta$ meteoroids \cite{wehry1999}.

Noise tests performed regularly during the 4 year period showed 
a degradation in the noise sensitivity of the dust instrument 
which is caused by a reduction  in the channeltron amplification.
This, however, has no effect on the interpretation of the present
data set.

\vspace{1cm}

{\bf Acknowledgments.}
We thank the Ulysses project at ESA and NASA/JPL for effective and successful 
mission operations. Our referees, Alexander V. Krivov and 
Larry W. Esposito, provided valuable suggestions which improved the 
presentation of our results.
This work has been supported by the Deutsches Zentrum f\"ur 
Luft- und Raumfahrt e.V. (DLR). 

\clearpage

\section*{Appendix}

{\normalsize
\begin{table}[htb]
%\caption{\label{mass_size} 
Conversion of impact charge, particle mass and particle radius for 
dust grains of different sizes and impact speeds.  Assuming spherical 
homogeneous particles with a
density of $\rm \rho = 2\, g\,cm^{-3}$, the grain radius 
can be directly calculated from the grain mass ($m=4\pi/3\cdot \rho \cdot r^3$,
column 2). 
The impact charge created when a dust particle with the mass given in column (1)
hits the detector target (columns 3 and 4) has been derived 
from the instrument calibration published by Gr\"un et al. 
(1995), their Fig. 3a).
The conversion between particle mass and ion charge amplitude range AR 
(columns 5 and 6) 
has been taken from Fig. 4 of the same publication.
The impact charge and the corresponding amplitude 
range depend on the particle's impact speed. Values 
are given for two different speeds (5 and $\rm 23\,km\,sec^{-1}$, respectively). 
It should be noted that the Ulysses and Galileo dust instruments could
only be calibrated in the laboratory up to masses of $\rm 10^{-12}\,g$ at impact 
speeds of $\rm 10\,km\,sec^{-1}$ and up to $\rm 10^{-10}\,g$ at impact speeds of 
$\rm 2\,km\,sec^{-1}$, respectively. For larger masses the calibration had 
to be extrapolated (Gr\"un~et~al.~1992a). The validity of this 
extrapolation, however, is supported by calibration measurements for 
other dust instruments (G\"oller~et~al.~1987).
%}
  \begin{tabular}{cccccc}
   \hline
   \hline \\[-2.0ex]
%Titelzeile
Mass & 
Radius &  
\multicolumn{2}{c}{Impact Charge $Q_I$} &  
\multicolumn{2}{c}{Amplitude Range} \\
    &
    &
{\footnotesize $\rm v =5\,km\,sec^{-1}$} &
{\footnotesize $\rm v = 23\,km\,sec^{-1}$} &
{\footnotesize $\rm v = 5\,km\,sec^{-1}$} &
{\footnotesize $\rm v = 23\,km\,sec^{-1}$} \\
(g) & 
[$\rm \mu m$] &
[C] &
[C] &
    &
    \\
(1) &
(2) &
(3) &
(4) &
(5) &
(6) \\
[0.7ex]
\hline \\[-2.0ex]
$1.2 \cdot 10^{-14}$ & 0.1 & $3.6 \cdot 10^{-16}$ & $3.2 \cdot 10^{-14}$ & --  & AR1 \\
$1.7 \cdot 10^{-13}$ & 0.2 & $5.0 \cdot 10^{-15}$ & $4.5 \cdot 10^{-13}$ & --  & AR2 \\
$1.9 \cdot 10^{-12}$ & 0.6 & $6.0 \cdot 10^{-14}$ & $5.0 \cdot 10^{-12}$ & AR1 & AR3 \\
$1.7 \cdot 10^{-11}$ & 1.2 & $5.1 \cdot 10^{-13}$ & $4.6 \cdot 10^{-11}$ & AR2 & AR4 \\
$1.5 \cdot 10^{-10}$ & 2.6 & $4.4 \cdot 10^{-12}$ & $4.0 \cdot 10^{-10}$ & AR3 & AR5 \\ [0.7ex] 
\hline
\end{tabular}\\[1.5ex]
\end{table}
}
\nocite{gruen1995a}
\nocite{goeller1987}

\hspace{2cm}

\clearpage

\bibliography{/home/krueger/tex/bib/pape,/home/krueger/tex/bib/references}

\clearpage

\section*{Tables}

\begin{table}[h]
\caption{\label{orbital_elements}
Numerical values of the orbital elements for the Ulysses 
trajectory (http://helio.estec.esa.nl/Ulysses/).
The reference frame is the mean ecliptic and equinox of 1950.0.
$a$ is the semi-major axis, $i$ the inclination, $e$ the eccentricity, 
$\Omega$ the longitude of ascending node  measured from the 
vernal equinox, $w$ the argument of perihelion and $M$ the mean anomaly
of the orbit. $J$ is the Julian date of interest.
}
\begin{tabular}{ccc}
\hline
$a$ & = &    3.3730032 AU (504594094 km) \\
$e$ & = &     0.60306 \\
$i$ & = &     79.12801$^{\circ}$ \\
$\Omega$ & = & -22.51862$^{\circ}$ \\
$w$ & = &     -1.11377$^{\circ}$ \\
$M$ & = &     0.1591096*(J - 2449788.986)$^{\circ}$ \\
\hline
\end{tabular}
\end{table}

{\small
\begin{table}[htb]
\caption{\label{event_table} Ulysses mission and dust detector (GRU) 
configuration, noise tests and other events. 
%GRU nominal configuration: EVD=C,I, SSEN=0,0,0,1(P), HV=3. 
%GRU noise test:  EVD=C,I,E; SSEN=0000; EVD=C,I; HV=4; nominal config.
Only selected events are given before 1996. See Sect.~\ref{mission} for 
details.
}
\small
  \begin{tabular*}{15.2cm}{lccl}
   \hline
   \hline \\[-2.0ex]
%Titelzeile
Yr-day &  
Date & 
Time &  
Event \\[0.7ex]
\hline \\[-2.0ex]
90-279 & 06.10.90 &       & Ulysses launch \\
92-039 & 08.02.92 & 12:04 & Ulysses Jupiter closest approach \\
95-300 & 27.10.95 & 16:23 & GRU heater: 1,200~mW \\
95-346 & 12.12.95 & 08:54 & GRU nominal configuration: HV=3, EVD=C,I, SSEN=0001 \\
96-011 & 11.01.96 & 05:59 & GRU noise test \\
96-040 & 09.02.96 & 05:00 & GRU noise test \\
96-067 & 07.03.96 & 21:00 & GRU noise test \\
96-096 & 05.04.96 & 20:00 & GRU noise test \\
96-123 & 02.05.96 & 15:59 & GRU noise test \\
96-151 & 30.05.96 & 20:59 & GRU noise test \\
96-180 & 28.06.96 & 13:00 & GRU noise test \\
96-207 & 25.07.96 & 10:00 & GRU noise test \\
96-230 & 17.08.96 & 16:00 & Ulysses DNEL \# 7 \\
96-231 & 18.08.96 & 01:28 & GRU on    \\
96-233 & 20.08.96 & 01:02 & GRU nominal configuration \\ 
96-237 & 24.08.96 & 09:00 & GRU noise test \\
96-263 & 19.09.96 & 16:59 & GRU noise test \\
96-291 & 17.10.96 & 05:59 & GRU noise test \\
96-319 & 14.11.96 & 11:59 & GRU noise test \\
96-347 & 12.12.96 & 03:00 & GRU noise test \\
97-009 & 09.01.97 & 05:59 & GRU noise test \\
97-037 & 05.02.97 & 00:00 & GRU noise test \\
97-065 & 06.03.97 & 01:00 & GRU noise test \\
97-091 & 01.04.97 & 20:55 & Ulysses DNEL \# 8 \\
97-092 & 02.04.97 & 05:21 & GRU on    \\
97-093 & 03.04.97 & 03:29 & GRU nominal configuration \\ 
97-093 & 03.04.97 & 18:59 & GRU noise test \\
97-121 & 01.05.97 & 03:00 & GRU noise test \\
97-149 & 29.05.97 & 00:00 & GRU noise test \\
97-177 & 26.05.97 & 00:00 & GRU noise test \\
97-205 & 24.07.97 & 17:00 & GRU noise test \\
97-233 & 21.08.97 & 08:00 & GRU noise test \\
97-261 & 18.09.97 & 06:00 & GRU noise test \\
97-290 & 17.10.97 & 04:00 & GRU noise test \\
97-317 & 13.11.97 & 04:00 & GRU noise test \\
97-345 & 11.12.97 & 01:16 & GRU noise test \\
98-008 & 08.01.98 & 03:59 & GRU noise test \\
98-036 & 05.02.98 & 09:22 & GRU noise test \\
[0.7ex] 
\hline
\end{tabular*}\\[1.5ex]
Abbreviations used: DNEL: Disconnect non-essential loads (i.~e.
all scientific instruments); 
HV: channeltron high voltage step; EVD: event definition,
ion- (I), channeltron- (C), or electron-channel (E); SSEN: detection thresholds,
ICP, CCP, ECP and PCP
\end{table}
}

\addtocounter{table}{-1}

{\small
\begin{table}[htb]
\caption{
 continued.
}
\small
  \begin{tabular*}{15.2cm}{lccl}
   \hline
   \hline \\[-2.0ex]
%Titelzeile
Yr-day &  
Date & 
Time &  
Event \\[0.7ex]
\hline \\[-2.0ex]
98-064 & 05.03.98 & 05:00 & GRU noise test \\
98-092 & 02.04.98 & 02:00 & GRU noise test \\
98-110 & 20.04.98 &       & Ulysses aphelion passage (5.4 AU) \\
98-120 & 30.04.98 & 19:59 & GRU noise test \\
98-129 & 09.05.98 &       & Ulysses ecliptic plane crossing (5.4 AU) \\
98-148 & 28.05.98 & 17:59 & GRU noise test \\
98-176 & 25.06.98 & 17:00 & GRU noise test \\
98-204 & 23.07.98 & 10:00 & GRU noise test \\
98-232 & 20.08.98 & 08:00 & GRU noise test \\
98-260 & 17.09.98 & 00:00 & GRU noise test \\
98-288 & 15.10.98 & 00:00 & GRU noise test \\
98-318 & 14.11.98 & 08:00 & GRU noise test \\
98-344 & 10.12.98 & 16:00 & GRU noise test \\
99-007 & 07.01.99 & 16:30 & GRU noise test \\
99-035 & 04.02.99 & 13:00 & GRU noise test \\
99-046 & 15.02.99 & 10:45 & Ulysses DNEL \# 9 \\
99-047 & 16.02.99 & 00:01 & GRU on    \\
99-048 & 17.02.99 & 04:29 & GRU nominal configuration \\ 
99-064 & 05.03.99 & 14:00 & GRU noise test \\
99-092 & 02.04.99 & 12:00 & GRU noise test \\
99-119 & 29.04.99 & 10:00 & GRU noise test \\
99-147 & 27.05.99 & 06:00 & GRU noise test \\
99-175 & 24.06.99 & 06:00 & GRU noise test \\
99-203 & 22.07.99 & 01:00 & GRU noise test \\
99-232 & 20.08.99 & 03:00 & GRU noise test \\
99-260 & 17.09.99 & 00:00 & GRU noise test \\
99-287 & 14.10.99 & 01:00 & GRU noise test \\
99-315 & 11.11.99 & 17:00 & GRU noise test \\
99-343 & 09.12.99 & 16:00 & GRU noise test \\
[0.7ex] 
\hline
\end{tabular*}\\[1.5ex]
\end{table}
}
\clearpage

\thispagestyle{empty}
\begin{sidewaystable}
\tabcolsep1.5mm
\tiny
\vbox{
\hspace{-2cm}
\begin{minipage}[t]{22cm}   
{\bf Table 3.} Overview of dust impacts detected with the Ulysses dust detector
between 1 January 1996 and 31 December 1999 as derived from the 
accumulators$^{\ast}$.  Switch-on of the instrument  
is indicated by horizontal lines. 
The heliocentric distance R, 
the lengths of the time interval $\Delta $t (days) from the previous 
table entry, and the corresponding numbers of impacts are given for the 
24 accumulators. The accumulators are arranged with increasing signal 
amplitude ranges (AR), with four event classes for each amplitude 
range (CLN = 0,1,2,3); e.g.~AC31 means counter for AR = 1 and CLN = 3.  
The $\Delta $t in the first line (96-001) 
is the time interval counted from the last entry in Table~2 in 
Paper~V. The totals of  counted impacts$^{\ast}$, of impacts with 
complete data, and of all events (noise plus impact events) 
for the entire period are given as well. 
\end{minipage}
}
\bigskip
\hspace{-2cm} 
\begin{tabular}{|r|r|r|r|cccc|cccc|cccc|cccc|cccc|cccc|}
\hline
&&&&& &&&&& &&&&& &&&&& &&&&& &&\\
\mc{1}{|c|}{Date}&
\mc{1}{|c|}{Time}&
\mc{1}{|c|}{R}&
\mc{1}{|c|}{$\Delta $t}&
%\mc{4}{|c|}{AR-1}&
%\mc{4}{|c|}{AR-2}&
%\mc{4}{|c|}{AR-3}&
%\mc{4}{|c|}{AR-4}&
%\mc{4}{|c|}{AR-5}&
%\mc{4}{|c|}{AR-6}\\
{\scriptsize AC}&{\scriptsize AC}&{\scriptsize AC}&{\scriptsize AC}&
{\scriptsize AC}&{\scriptsize AC}&{\scriptsize AC}&{\scriptsize AC}&
{\scriptsize AC}&{\scriptsize AC}&{\scriptsize AC}&{\scriptsize AC}&
{\scriptsize AC}&{\scriptsize AC}&{\scriptsize AC}&{\scriptsize AC}&
{\scriptsize AC}&{\scriptsize AC}&{\scriptsize AC}&{\scriptsize AC}&
{\scriptsize AC}&{\scriptsize AC}&{\scriptsize AC}&{\scriptsize AC}\\
&&[AU]&\mc{1}{c|}{[d]}&
\,\,01$^\ast$&\,\,11$^\ast$&21&31&
\,\,02$^\ast$&12&22&32&
03&13&23&33&
04&14&24&34&
05&15&25&35&
06&16&26&36\\
&&&&& &&&&& &&&&& &&&&& &&&&& &&\\
\hline
&&&&& &&&&& &&&&& &&&&& &&&&& &&\\
96-001&20:55&      3.060& 5.948&-&-&-&-&-&-&-&-&-&-&-&1&-&-&-&-&-&-&-&-&-&-&-&-\\
96-015&21:10&      3.150& 14.01&2&-&-&-&-&-&-&-&1&-&-&2&-&-&-&-&-&-&-&-&-&-&-&-\\
96-032&07:20&      3.240& 16.42&3&-&-&-&-&-&-&1&-&-&-&1&-&-&-&-&-&-&-&-&-&-&-&-\\
96-049&19:34&      3.350& 17.50&3&-&-&-&-&-&-&-&-&-&-&-&-&-&-&-&-&-&-&-&-&-&-&-\\
96-067&22:35&      3.450& 18.12&4&-&-&-&-&1&-&1&-&-&-&1&-&-&-&-&-&-&-&-&-&-&-&-\\
 &&&&& &&&&& &&&&& &&&&& &&&&& &&\\
96-082&00:44&      3.520& 14.08&-&1&-&-&-&-&-&-&-&-&-&2&-&-&-&-&-&-&-&-&-&-&-&-\\
96-096&12:00&      3.600& 14.46&2&-&-&-&-&-&-&-&-&-&-&1&-&-&-&1&-&-&-&-&-&-&-&-\\
96-119&10:23&      3.720& 22.93&3&-&-&1&1&-&-&1&-&-&-&2&-&-&-&-&-&-&-&-&-&-&-&-\\
96-145&01:12&      3.850& 25.61&4&-&-&-&1&1&-&-&-&-&1&2&-&-&-&-&-&-&-&-&-&-&-&-\\
96-170&12:28&      3.970& 25.46&2&-&-&-&-&1&-&-&-&-&1&-&-&-&-&-&-&-&-&-&-&-&-&-\\
 &&&&& &&&&& &&&&& &&&&& &&&&& &&\\
96-194&15:47&      4.080& 24.13&2&-&-&-&-&-&-&-&-&-&-&2&-&-&-&-&-&-&-&-&-&-&-&-\\
96-221&07:41&      4.190& 26.66&1&-&-&-&1&-&-&-&-&-&-&-&-&-&-&-&-&-&-&-&-&-&-&-\\
96-230&05:20&      4.230& 8.902&-&-&-&-&-&-&-&-&-&-&-&-&-&-&-&-&-&-&-&-&-&-&-&-\\
96-233&01:02&      4.240& 2.820&\mc{4}{|c|}{---------------------}&\mc{4}{|c|}{---------------------}&\mc{4}{|c|}{---------------------}&\mc{4}{|c|}{---------------------}&\mc{4}{|c|}{---------------------}&\mc{4}{|c|}{---------------------}\\
 &&&&& &&&&& &&&&& &&&&& &&&&& &&\\
96-254&03:25&      4.320& 21.09&-&1&-&-&-&-&-&-&-&-&-&-&-&-&-&-&-&-&-&-&-&-&-&-\\
96-281&02:43&      4.420& 26.97&1&-&-&-&-&-&-&-&-&-&-&-&-&-&-&1&-&-&-&-&-&-&-&-\\
96-305&08:51&      4.510& 24.25&2&-&-&-&-&-&-&-&-&1&-&-&-&-&-&-&-&-&-&-&-&-&-&-\\
96-320&22:16&      4.570& 15.55&-&-&-&-&-&-&-&2&-&1&-&1&-&-&-&-&-&-&-&-&-&-&-&-\\
96-359&13:34&      4.690& 38.63&1&-&-&-&-&-&-&1&-&-&-&-&-&-&-&1&-&-&-&-&-&-&-&-\\
 &&&&& &&&&& &&&&& &&&&& &&&&& &&\\
97-015&13:30&      4.760& 20.99&-&-&-&-&-&-&-&-&-&1&-&1&-&-&-&-&-&-&-&-&-&-&-&-\\
97-028&01:22&      4.790& 12.49&1&-&-&-&-&-&-&-&-&-&-&-&-&-&-&-&-&-&-&-&-&-&-&-\\
97-060&20:57&      4.880& 32.81&-&-&-&-&-&-&-&1&-&-&-&1&-&-&-&-&-&-&-&-&-&-&-&-\\
97-091&17:50&      4.960& 30.87&-&-&-&-&1&1&-&-&-&-&-&-&-&-&-&-&-&-&-&-&-&-&-&-\\
97-093&03:29&      4.960& 1.402&\mc{4}{|c|}{---------------------}&\mc{4}{|c|}{---------------------}&\mc{4}{|c|}{---------------------}&\mc{4}{|c|}{---------------------}&\mc{4}{|c|}{---------------------}&\mc{4}{|c|}{---------------------}\\
97-111&23:30&      5.010& 18.83&-&-&-&-&1&-&-&-&-&-&-&-&-&-&-&-&-&-&-&-&-&-&-&-\\
97-145&15:34&      5.080& 33.66&-&-&-&-&-&-&-&-&1&-&-&1&-&-&-&1&-&-&-&-&-&-&-&-\\
97-170&02:14&      5.130& 24.44&3&-&-&-&-&1&-&-&-&-&-&2&-&-&-&-&-&-&-&-&-&-&-&-\\
97-223&08:01&      5.220& 53.24&-&-&-&-&1&-&-&1&-&-&-&-&-&-&-&-&-&-&-&-&-&-&-&-\\
 &&&&& &&&&& &&&&& &&&&& &&&&& &&\\
97-253&11:13&      5.260& 30.13&2&1&-&-&-&-&-&-&-&1&-&2&-&-&-&-&-&-&-&-&-&-&-&1\\
97-295&21:15&      5.320& 42.41&-&-&-&-&-&-&-&1&-&-&1&-&-&-&-&-&-&-&-&-&-&-&-&-\\
97-325&01:21&      5.350& 29.17&1&-&-&-&-&-&-&1&-&-&-&1&-&-&-&1&-&-&-&-&-&-&-&-\\
97-353&03:25&      5.370& 28.08&-&-&-&-&-&-&-&-&-&1&-&-&-&1&1&-&-&-&-&-&-&-&-&-\\
98-008&01:41&      5.380& 19.92&3&-&-&-&-&-&-&2&-&-&-&-&-&-&-&1&-&-&-&-&-&-&-&-\\
 &&&&& &&&&& &&&&& &&&&& &&&&& &&\\
\hline
\end{tabular}
\end{sidewaystable}

\clearpage
\thispagestyle{empty}
\begin{sidewaystable}
\tabcolsep1.5mm
\tiny
\hspace{-2cm}   
\begin{tabular}{|r|r|r|r|cccc|cccc|cccc|cccc|cccc|cccc|}
\hline
&&&&& &&&&& &&&&& &&&&& &&&&& &&\\
\mc{1}{|c|}{Date}&
\mc{1}{|c|}{Time}&
\mc{1}{|c|}{R}&
\mc{1}{|c|}{$\Delta $t}&
%\mc{4}{|c|}{AR-1}&
%\mc{4}{|c|}{AR-2}&
%\mc{4}{|c|}{AR-3}&
%\mc{4}{|c|}{AR-4}&
%\mc{4}{|c|}{AR-5}&
%\mc{4}{|c|}{AR-6}\\
{\scriptsize AC}&{\scriptsize AC}&{\scriptsize AC}&{\scriptsize AC}&
{\scriptsize AC}&{\scriptsize AC}&{\scriptsize AC}&{\scriptsize AC}&
{\scriptsize AC}&{\scriptsize AC}&{\scriptsize AC}&{\scriptsize AC}&
{\scriptsize AC}&{\scriptsize AC}&{\scriptsize AC}&{\scriptsize AC}&
{\scriptsize AC}&{\scriptsize AC}&{\scriptsize AC}&{\scriptsize AC}&
{\scriptsize AC}&{\scriptsize AC}&{\scriptsize AC}&{\scriptsize AC}\\
&&[AU]&\mc{1}{c|}{[d]}&
\,\,01$^\ast$&\,\,11$^\ast$&21&31&
\,\,02$^\ast$&12&22&32&
03&13&23&33&
04&14&24&34&
05&15&25&35&
06&16&26&36\\
&&&&& &&&&& &&&&& &&&&& &&&&& &&\\
\hline
&&&&& &&&&& &&&&& &&&&& &&&&& &&\\
98-036&00:05&      5.400& 27.93&2&-&-&-&-&-&-&1&-&-&-&1&-&-&-&-&-&-&-&-&-&-&-&-\\
98-069&04:14&      5.410& 33.17&1&-&-&-&-&-&-&-&-&1&-&-&-&-&-&-&-&-&-&-&-&-&-&-\\
98-080&22:21&      5.410& 11.75&-&-&-&-&-&-&-&-&-&-&-&1&-&-&-&-&-&-&-&-&-&-&-&-\\
98-102&15:22&      5.410& 21.70&3&2&-&-&-&1&-&-&-&-&-&2&-&-&-&-&-&-&-&-&-&-&-&-\\
98-139&17:34&      5.410& 37.09&-&2&-&-&-&-&-&-&-&1&-&-&-&-&-&-&-&-&-&-&-&-&-&-\\
 &&&&& &&&&& &&&&& &&&&& &&&&& &&\\
98-173&13:59&      5.400& 33.85&2&2&-&1&-&-&-&-&-&-&-&1&-&-&-&-&-&-&-&-&-&-&-&-\\
98-205&22:27&      5.380& 32.35&2&-&-&-&1&1&-&-&-&-&-&1&-&-&-&-&-&-&-&-&-&-&-&-\\
98-236&02:46&      5.360& 30.17&4&-&-&-&1&1&-&-&-&-&-&4&-&-&-&-&-&-&-&-&-&-&-&-\\
98-273&09:10&      5.330& 37.26&-&-&-&1&-&2&-&3&-&-&-&1&-&-&-&-&-&-&-&-&-&-&-&-\\
98-311&16:27&      5.280& 38.30&-&-&-&-&-&-&-&-&-&1&-&2&-&-&-&-&-&-&-&-&-&-&-&-\\
 &&&&& &&&&& &&&&& &&&&& &&&&& &&\\
98-350&21:24&      5.230& 39.20&1&-&-&-&-&-&-&1&-&-&-&-&-&-&-&-&-&-&-&-&-&-&-&-\\
98-363&17:30&      5.210& 12.83&-&-&-&-&-&-&-&-&-&-&-&-&-&-&-&-&-&1&-&-&-&-&-&-\\
99-016&08:38&      5.180& 17.63&2&-&-&-&1&1&-&1&-&-&-&-&-&-&-&-&-&-&-&-&-&-&-&-\\
99-046&09:00&      5.130& 30.01&1&-&-&-&-&-&-&-&-&-&-&-&-&-&-&-&-&-&-&-&-&-&-&-\\
99-048&04:29&      5.120& 1.811&\mc{4}{|c|}{---------------------}&\mc{4}{|c|}{---------------------}&\mc{4}{|c|}{---------------------}&\mc{4}{|c|}{---------------------}&\mc{4}{|c|}{---------------------}&\mc{4}{|c|}{---------------------}\\
99-058&06:44&      5.100& 10.09&1&1&-&-&-&-&-&-&-&-&-&-&-&-&-&-&-&-&-&-&-&-&-&-\\
99-088&15:36&      5.040& 30.36&2&-&-&-&-&1&-&-&-&-&1&2&-&-&-&-&-&-&-&-&-&-&-&-\\
99-112&20:59&      4.990& 24.22&2&-&-&-&-&-&-&-&-&-&-&-&-&-&-&-&-&-&-&-&-&-&-&-\\
99-137&16:01&      4.930& 24.79&-&-&-&1&1&-&-&1&-&-&-&-&-&-&-&-&-&-&-&-&-&-&-&-\\
 &&&&& &&&&& &&&&& &&&&& &&&&& &&\\
99-150&20:38&      4.890& 13.19&-&-&-&-&1&-&-&-&-&-&-&-&-&-&-&-&-&-&-&-&-&-&-&-\\
99-181&00:58&      4.810& 30.18&1&-&-&-&3&-&-&1&-&-&-&-&-&-&-&-&-&-&-&-&-&-&-&-\\
99-198&18:23&      4.760& 17.72&1&-&-&-&-&1&-&-&-&-&-&-&-&-&-&-&-&-&-&-&-&-&-&-\\
99-218&18:45&      4.700& 20.01&-&-&-&-&-&-&-&-&-&-&-&1&-&-&-&-&-&-&-&-&-&-&-&-\\
99-242&11:53&      4.620& 23.71&3&-&-&-&-&1&-&-&-&1&-&1&-&-&-&-&-&-&-&-&-&-&-&-\\
 &&&&& &&&&& &&&&& &&&&& &&&&& &&\\
99-272&21:01&      4.520& 30.38&1&-&-&1&1&-&-&-&-&-&-&-&-&-&-&1&-&-&-&-&-&-&-&-\\
99-286&02:43&      4.480& 13.23&2&-&-&-&1&-&-&1&-&-&-&1&-&-&-&-&-&-&-&-&-&-&-&-\\
99-308&21:21&      4.390& 22.77&1&-&-&-&-&1&-&-&-&-&-&-&-&-&-&-&-&-&-&-&-&-&-&-\\
99-333&04:30&      4.300& 24.29&3&-&-&-&-&-&-&-&-&-&-&1&-&-&-&1&-&-&-&-&-&-&-&-\\
99-360&04:15&      4.190& 26.98&1&1&-&-&2&1&-&-&-&-&-&1&-&-&-&1&-&-&-&-&-&-&-&-\\
 &&&&& &&&&& &&&&& &&&&& &&&&& &&\\
\hline 
\mc{4}{|l|}{}&& &&&&& &&&&& &&&&& &&&&& &&\\
\mc{4}{|l|}{Impacts (counted)}&-- &-- &0&5&-- &16&0&21&2&9&4&43&0&1&1&9&0&1&0&0&0&0&0&1\\[1.5ex]
\mc{4}{|l|}{Impacts (complete data)}&76&11&0&5&18&16&0&21&2&9&4&43&0&1&1&9&0&1&0&0&0&0&0&1\\[1.5ex]
\mc{4}{|l|}{All events(complete data)}&15295&1033&0&5&312&16&0&21&7&9&4&43&0&1&1&9&0&1&0&0&0&0&0&1\\[1.5ex]
\hline
\end{tabular}
\\
\parbox{19cm}{
$\ast$: {\small Entries for AC01, AC11 and AC02 are the number of impacts 
with complete data. Due to the noise contamination of these three categories 
the number of impacts cannot be determined from the accumulators. The method 
to separate dust impacts from noise events in these 
three categories has been given 
by Baguhl et al. (1993)}
}
\end{sidewaystable}

\clearpage

\thispagestyle{empty}
 \begin{sidewaystable}
\tabcolsep1.2mm
\tiny
\vbox{
\hspace{-2cm}
\begin{minipage}[t]{21.5cm}      
 {\bf Table 4.}
 Raw data: No., impact time, CLN, AR, SEC, IA, EA, CA, IT, ET, 
 EIT, EIC, ICC, PA, PET, EVD, ICP, ECP, CCP, PCP, HV and 
 evaluated data: R, LON, LAT, $D_{\rm Jup}$ (in astronomical units), rotation angle 
(ROT), instr. pointing (S$_{LON}$, S$_{LAT}$), impact speed ($v$, in $\rm km\,sec^{-1}$), 
 speed error factor (VEF), mass ($m$, in grams) and mass error factor (MEF).
\end{minipage}
}
\bigskip

\hspace{-2cm}                                                                    
\begin{tabular*}{23cm} { %@{\extracolsep{\fill}}                                  
rrrrrrrrrrrrrrrcrrrrcrrrrrrrrrrrrrrrrrrrrr}                                     
 \mc{1}{c}{No.}&               \mc{1}{c}{IMP. DATE}&         \mc{1}{c}{C}&                 \mc{1}{c}{AR}&                
 \mc{1}{c}{S}&                 \mc{1}{c}{IA}&                \mc{1}{c}{EA}&                \mc{1}{c}{CA}&                
 \mc{1}{c}{IT}&                \mc{1}{c}{ET}&                \mc{1}{c}{E}&                 \mc{1}{c}{E}&                 
 \mc{1}{c}{I}&                 \mc{1}{c}{PA}&                \mc{1}{c}{P}&                 \mc{1}{c}{E}&                 
 \mc{1}{c}{I}&                 \mc{1}{c}{E}&                 \mc{1}{c}{C}&                 \mc{1}{c}{P}&                 
 \mc{1}{c}{HV}&                \mc{1}{c}{R}&                 \mc{1}{c}{LON}&               \mc{1}{c}{LAT}&               
 \mc{1}{c}{D$_{\rm Jup}$}&     \mc{1}{c}{ROT}&               \mc{1}{c}{S$_{\rm LON}$}&     \mc{1}{c}{S$_{\rm LAT}$}&     
 \mc{1}{c}{V}&                 \mc{1}{c}{VEF}&               \mc{1}{c}{M}&                 \mc{1}{c}{MEF}\\              
 &                             &                             \mc{1}{c}{L}&                 &                             
 \mc{1}{c}{E}&                 &                             &                             &                             
 &                             &                             \mc{1}{c}{I}&                 \mc{1}{c}{I}&                 
 \mc{1}{c}{I}&                 &                             \mc{1}{c}{E}&                 \mc{1}{c}{V}&                 
 \mc{1}{c}{C}&                 \mc{1}{c}{C}&                 \mc{1}{c}{C}&                 \mc{1}{c}{C}&                 
 &                             &                             &                             &                             
 &                             &                             &                             &                             
 &                             &                             &                             \\                            
 &                             &                             \mc{1}{c}{N}&                 &                             
 \mc{1}{c}{C}&                 &                             &                             &                             
 &                             &                             \mc{1}{c}{T}&                 \mc{1}{c}{C}&                 
 \mc{1}{c}{C}&                 &                             \mc{1}{c}{T}&                 \mc{1}{c}{D}&                 
 \mc{1}{c}{P}&                 \mc{1}{c}{P}&                 \mc{1}{c}{P}&                 \mc{1}{c}{P}&                 
 &                             &                             &                             &                             
 &                             &                             &                             &                             
 &                             &                             &                             \\                            
&&&&& &&&&& &&&&& &&&&& &&&&& &&&&& &\\                                         
\hline                                                                          
&&&&& &&&&& &&&&& &&&&& &&&&& &&&&& &\\                                         
 1478&96-001 20:55&3&3&167&23&27&11& 8& 7& 6&0&1&47& 0&1&0&0&0&1&3&
3.05883&138.4& 59.6& 6.86682&114&246&-12&18.0& 1.6&$  7.6\cdot 10^{-12}$&   6.0\\
 1479&96-002 16:12&0&1&208& 3& 7& 0& 9& 8& 9&0&0& 0& 0&1&0&0&0&1&3&
3.06421&138.5& 59.5& 6.87260&171&187&-24&22.7& 1.6&$  7.8\cdot 10^{-15}$&   6.0\\
 1480&96-004 12:16&0&1&186& 5&10& 0& 7& 8& 8&0&0& 0& 0&1&0&0&0&1&3&
3.07571&138.7& 59.2& 6.88499&136&222&-18&35.4& 1.6&$  4.1\cdot 10^{-15}$&   6.0\\
 1481&96-006 11:13&3&3&199&19&24&15& 6& 6& 6&0&1&43& 0&1&0&0&0&1&3&
3.08719&138.9& 59.0& 6.89741&153&204&-22&28.0& 1.6&$  3.8\cdot 10^{-13}$&   6.0\\
 1482&96-009 04:15&0&3&220&19&22& 0& 7& 6& 6&0&0&41& 0&1&0&0&0&1&3&
3.10395&139.2& 58.6& 6.91566&177&176&-23&22.7& 1.6&$  5.8\cdot 10^{-13}$&   6.0\\
&&&&& &&&&& &&&&& &&&&& &&&&& &&&&& &\\                      \hline                                                      
 1483&96-015 21:10&3&3&177&22&28& 6& 6& 4& 5&0&1&47& 0&1&0&0&0&1&3&
3.14480&139.8& 57.6& 6.96062&103&246& -8&38.7& 1.6&$  4.4\cdot 10^{-13}$&   6.0\\
 1484&96-016 15:18&0&1&155& 4& 8& 0&10&15& 0&1&0&46&31&1&0&0&0&1&3&
3.14932&139.9& 57.5& 6.96563& 72&276&  1& 4.9& 1.7&$  1.3\cdot 10^{-12}$&   7.7\\
 1485&96-023 14:47&3&3&243&19&23& 2& 6& 6& 6&0&1&43& 0&1&0&0&0&1&3&
3.19043&140.5& 56.6& 7.01168&185&157&-22&28.0& 1.6&$  3.3\cdot 10^{-13}$&   6.0\\
 1486&96-024 21:53&0&1&182& 5& 2& 0& 7&15&15&1&0& 5&24&1&0&0&0&1&3&
3.19860&140.7& 56.4& 7.02092& 93&248& -5&34.1& 1.9&$  1.3\cdot 10^{-15}$&  10.5\\
 1487&96-026 00:11&3&2&120& 9&13& 7& 7& 8& 6&0&1&35& 0&1&0&0&0&1&3&
3.20527&140.7& 56.3& 7.02848&  6&333& 13&35.4& 1.6&$  1.3\cdot 10^{-14}$&   6.0\\
&&&&& &&&&& &&&&& &&&&& &&&&& &&&&& &\\                      \hline                                                      
 1488&96-032 07:20&0&1&128& 6&11& 0& 7& 7& 7&0&0& 6&31&1&0&0&0&1&3&
3.24210&141.3& 55.5& 7.07058&  4&329& 13&43.7& 1.6&$  2.3\cdot 10^{-15}$&   6.0\\
 1489&96-041 08:11&0&1&  0& 5&11& 0& 8& 7& 8&0&0& 2&24&1&0&0&0&1&3&
3.29450&142.0& 54.3& 7.13144&166&160&-23&34.6& 1.6&$  5.3\cdot 10^{-15}$&   6.0\\
 1490&96-042 09:46&0&1&213& 6& 9& 0& 9&15& 0&1&0& 7&31&1&0&0&0&1&3&
3.30099&142.0& 54.2& 7.13906&100&224& -7& 5.7& 2.1&$  1.3\cdot 10^{-12}$&  15.7\\
 1491&96-049 19:34&0&1&246& 4& 7& 0& 8& 9& 8&0&0& 0& 0&1&0&0&0&1&3&
3.34327&142.6& 53.3& 7.18913&134&186&-18&25.9& 1.6&$  5.7\cdot 10^{-15}$&   6.0\\
 1492&96-052 08:09&0&1&158& 1& 9& 0&15& 9&15&1&0&42&31&1&0&0&0&1&3&
3.35749&142.7& 53.0& 7.20613&  5&309& 16&11.8&11.8&$  7.3\cdot 10^{-14}$&5858.3\\
&&&&& &&&&& &&&&& &&&&& &&&&& &&&&& &\\                      \hline                                                      
 1493&96-058 20:58&0&1& 66& 1& 4& 0&15& 8&12&0&0& 0& 0&1&0&0&0&1&3&
3.39417&143.2& 52.2& 7.25039&225& 82&-20&11.8&11.8&$  3.0\cdot 10^{-14}$&5858.3\\
 1494&96-062 13:19&1&2& 39& 9& 4& 5& 8&12& 0&1&1& 0& 0&1&0&0&0&1&3&
3.41515&143.4& 51.7& 7.27596&182&127&-28&21.4& 1.9&$  1.6\cdot 10^{-14}$&  10.5\\
 1495&96-064 12:54&0&1& 97& 2& 5& 0&15&10&11&0&0& 0& 0&1&0&0&0&1&3&
3.42628&143.5& 51.5& 7.28961&259& 46& -8&26.5& 2.0&$  2.7\cdot 10^{-15}$&  12.5\\
 1496&96-064 14:18&3&2&  9&14&23& 8& 8& 7& 6&0&1&42& 0&1&0&0&0&1&3&
3.42628&143.5& 51.5& 7.28961&135&177&-21&21.4& 1.9&$  5.8\cdot 10^{-13}$&  10.5\\
 1497&96-067 01:32&0&1&116& 6& 3& 0& 8&15& 0&1&0&35&30&1&0&0&0&1&3&
3.44015&143.7& 51.2& 7.30667&281& 25&  0&21.4& 1.9&$  8.5\cdot 10^{-15}$&  10.5\\
&&&&& &&&&& &&&&& &&&&& &&&&& &&&&& &\\                      \hline                                                      
 1498&96-067 22:35&3&3&246&20&25& 6& 5& 6& 5&0&1&43& 0&0&0&0&0&0&3&
3.44499&143.7& 51.1& 7.31264&104&206& -9&35.4& 1.6&$  2.8\cdot 10^{-13}$&   6.0\\
 1499&96-072 07:45&1&1&208& 7& 5& 6&10&15& 0&1&1&38&31&1&0&0&0&1&3&
3.46908&144.0& 50.6& 7.34252& 42&261& 16&10.4& 1.9&$  1.3\cdot 10^{-13}$&  10.5\\
 1500&96-075 23:41&3&3&254&20&25& 1& 6& 6& 6&0&1&44& 0&1&0&0&0&1&3&
3.48890&144.2& 50.2& 7.36729&103&205&-10&28.0& 1.6&$  5.9\cdot 10^{-13}$&   6.0\\
 1501&96-082 00:43&3&3& 29&20&25& 5& 5& 5& 5&0&1&44& 0&1&0&0&0&1&3&
3.52212&144.5& 49.5& 7.40915&139&170&-27&43.7& 1.6&$  1.1\cdot 10^{-13}$&   6.0\\
 1502&96-088 06:00&3&3&242&22&28& 6& 7& 5& 6&0&1&47& 0&1&0&0&0&1&3&
3.55565&144.8& 48.8& 7.45101& 72&230&  5&28.1& 1.6&$  1.4\cdot 10^{-12}$&   6.0\\
&&&&& &&&&& &&&&& &&&&& &&&&& &&&&& &\\                      \hline                                                      
 1503&96-096 04:55&0&1& 46& 5& 3& 0& 7&15&15&1&0&19&30&1&0&0&0&1&3&
3.59739&145.2& 48.0& 7.50567&148&162&-34&34.1& 1.9&$  1.4\cdot 10^{-15}$&  10.5\\
 1504&96-096 08:40&0&1& 21& 5& 4& 0& 8&15& 0&1&0&37&30&1&0&0&0&1&3&
3.59804&145.2& 47.9& 7.50652&113&196&-16&21.4& 1.9&$  8.3\cdot 10^{-15}$&  10.5\\
 1505&96-096 12:00&3&4&237&25&49&18& 7& 4& 5&0&1&47& 0&1&0&0&0&1&3&
3.59870&145.2& 47.9& 7.50738& 56&243& 15&14.1& 1.9&$  6.5\cdot 10^{-11}$&  10.5\\
 1506&96-102 01:33&0&1&109& 5& 9& 0& 8& 7& 8&0&0& 5&24&1&0&0&0&1&3&
3.62815&145.5& 47.3& 7.54578&231& 61&-26&34.6& 1.6&$  3.8\cdot 10^{-15}$&   6.0\\
 1507&96-103 01:35&0&2&128& 9&14& 0& 7& 7& 6&0&0&37& 0&1&0&0&0&1&3&
3.63335&145.5& 47.2& 7.55261&258& 38&-10&43.7& 1.6&$  6.4\cdot 10^{-15}$&   6.0\\
&&&&& &&&&& &&&&& &&&&& &&&&& &&&&& &\\                      \hline                                                      
 1508&96-103 17:53&0&1& 45& 7&11& 0&11&15&15&1&0&99&31&1&0&0&0&1&3&
3.63660&145.5& 47.2& 7.55687&142&169&-33& 4.2& 1.6&$  5.7\cdot 10^{-12}$&   6.0\\
 1509&96-105 20:51&0&1&250& 4& 5& 0& 9&15& 0&1&0&17&31&1&0&0&0&1&3&
3.64762&145.6& 46.9& 7.57137& 68&233& 10&14.1& 1.9&$  3.8\cdot 10^{-14}$&  10.5\\
 1510&96-109 03:23&3&2& 35&14&21&13&10&11& 7&0&1&38& 0&1&0&0&0&1&3&
3.66503&145.8& 46.6& 7.59438&124&189&-24&10.4& 1.9&$  4.0\cdot 10^{-12}$&  10.5\\
 1511&96-109 16:46&3&1&209& 1& 4& 1&15&10&11&0&1& 0& 0&1&0&0&0&1&3&
3.66760&145.8& 46.5& 7.59779&  8&292& 35&11.8&11.8&$  3.0\cdot 10^{-14}$&5858.3\\
 1512&96-112 22:48&3&3& 40&19&23& 9& 7& 7& 6&0&1&41& 0&1&0&0&0&1&3&
3.68426&145.9& 46.2& 7.61993&129&184&-27&21.0& 1.6&$  9.6\cdot 10^{-13}$&   6.0\\
&&&&& &&&&& &&&&& &&&&& &&&&& &&&&& &\\                      \hline                                                      
 1513&96-119 10:23&3&3& 30&22&28&13& 6& 5& 5&0&1&47& 0&1&0&0&0&1&3&
3.71728&146.2& 45.5& 7.66416&111&201&-17&34.6& 1.6&$  7.0\cdot 10^{-13}$&   6.0\\
 1514&96-125 06:28&0&1& 27& 4& 6& 0&10&15& 0&1&0&22&31&1&0&0&0&1&3&
3.74680&146.4& 45.0& 7.70406&104&208&-12& 4.9& 1.7&$  9.0\cdot 10^{-13}$&   7.7\\
 1515&96-125 14:27&0&1&143& 1&10& 0&15& 9&15&1&0&43&31&1&0&0&0&1&3&
3.74805&146.4& 44.9& 7.70576&267& 33& -5&11.8&11.8&$  8.6\cdot 10^{-14}$&5858.3\\
 1516&96-126 11:55&0&2&202& 8&12& 0& 7& 7& 6&0&0&35& 0&1&0&0&0&1&3&
3.75242&146.5& 44.8& 7.71170&347&321& 39&43.7& 1.6&$  4.0\cdot 10^{-15}$&   6.0\\
 1517&96-130 23:59&3&3& 48&21&27& 3& 6& 5& 6&0&1&46& 0&1&0&0&0&1&3&
3.77477&146.6& 44.4& 7.74219&130&189&-30&34.6& 1.6&$  5.0\cdot 10^{-13}$&   6.0\\
&&&&& &&&&& &&&&& &&&&& &&&&& &&&&& &\\                      \hline                                                      
 1518&96-133 11:14&1&2& 68&14&14& 7& 8& 8& 0&1&1&39& 2&1&0&0&0&1&3&
3.78711&146.7& 44.2& 7.75912&156&161&-45&21.4& 1.9&$  2.0\cdot 10^{-13}$&  10.5\\
 1519&96-135 17:11&0&1&204& 3& 8& 0& 8& 7& 9&0&0& 0& 0&1&0&0&0&1&3&
3.79817&146.8& 43.9& 7.77433&345&326& 41&34.6& 1.6&$  2.3\cdot 10^{-15}$&   6.0\\
 1520&96-139 02:21&3&3& 10&22&28& 9& 8& 6& 6&0&1&46& 0&1&0&0&0&1&3&
3.81467&146.9& 43.6& 7.79713& 71&233& 10&19.5& 1.6&$  5.6\cdot 10^{-12}$&   6.0\\
 1521&96-140 06:31&2&3& 57&19&23&22& 6& 5& 6&0&1&42& 0&1&0&0&0&1&3&
3.82076&147.0& 43.5& 7.80556&138&184&-37&34.6& 1.6&$  1.7\cdot 10^{-13}$&   6.0\\
 1522&96-145 01:12&0&1&231& 3& 8& 0& 8& 8&10&0&0& 2&24&1&0&0&0&1&3&
3.84376&147.1& 43.0& 7.83757& 19&282& 41&28.0& 1.6&$  4.5\cdot 10^{-15}$&   6.0\\
&&&&& &&&&& &&&&& &&&&& &&&&& &&&&& &\\                      \hline                                                      
 1523&96-154 21:09&1&2& 30&13& 9&10&10&15& 0&1&1&36&31&1&0&0&0&1&3&
3.89094&147.5& 42.1& 7.90387& 92&222& -4&10.4& 1.9&$  7.2\cdot 10^{-13}$&  10.5\\
 1524&96-164 10:49&2&3& 24&21&26& 8& 6& 5& 6&0&1&45& 0&1&0&0&0&1&3&
3.93553&147.8& 41.3& 7.96733& 80&232&  5&34.6& 1.6&$  4.2\cdot 10^{-13}$&   6.0\\
 1525&96-169 04:12&0&1&  5& 4& 3& 0&10&15& 0&1&0&39&31&1&0&0&0&1&3&
3.95753&147.9& 40.8& 7.99892& 50&254& 28&10.4& 1.9&$  5.8\cdot 10^{-14}$&  10.5\\
 1526&96-170 12:28&0&1& 19& 3& 2& 0&10&15& 0&1&0&36&31&1&0&0&0&1&3&
3.96386&148.0& 40.7& 8.00804& 70&239& 13&10.4& 1.9&$  4.4\cdot 10^{-14}$&  10.5\\
 1527&96-175 20:50&0&1&139& 4& 5& 0& 7&15& 0&1&0&42&31&1&0&0&0&1&3&
3.98788&148.1& 40.2& 8.04281&236& 62&-30&34.1& 1.9&$  1.6\cdot 10^{-15}$&  10.5\\
\end{tabular*} \end{sidewaystable}                                                     
\clearpage                                                                      
\thispagestyle{empty}
\begin{sidewaystable}   
\tiny \hspace{-2cm}         
\tabcolsep1.2mm
\begin{tabular*}{23cm} {%@{\extracolsep{\fill}}                                  
rrrrrrrrrrrrrrrcrrrrcrrrrrrrrrrrrrrrrrrrrr}                                     
 \mc{1}{c}{No.}&               \mc{1}{c}{IMP. DATE}&         \mc{1}{c}{C}&                 \mc{1}{c}{AR}&                
 \mc{1}{c}{S}&                 \mc{1}{c}{IA}&                \mc{1}{c}{EA}&                \mc{1}{c}{CA}&                
 \mc{1}{c}{IT}&                \mc{1}{c}{ET}&                \mc{1}{c}{E}&                 \mc{1}{c}{E}&                 
 \mc{1}{c}{I}&                 \mc{1}{c}{PA}&                \mc{1}{c}{P}&                 \mc{1}{c}{E}&                 
 \mc{1}{c}{I}&                 \mc{1}{c}{E}&                 \mc{1}{c}{C}&                 \mc{1}{c}{P}&                 
 \mc{1}{c}{HV}&                \mc{1}{c}{R}&                 \mc{1}{c}{LON}&               \mc{1}{c}{LAT}&               
 \mc{1}{c}{D$_{\rm Jup}$}&     \mc{1}{c}{ROT}&               \mc{1}{c}{S$_{\rm LON}$}&     \mc{1}{c}{S$_{\rm LAT}$}&     
 \mc{1}{c}{V}&                 \mc{1}{c}{VEF}&               \mc{1}{c}{M}&                 \mc{1}{c}{MEF}\\              
 &                             &                             \mc{1}{c}{L}&                 &                             
 \mc{1}{c}{E}&                 &                             &                             &                             
 &                             &                             \mc{1}{c}{I}&                 \mc{1}{c}{I}&                 
 \mc{1}{c}{I}&                 &                             \mc{1}{c}{E}&                 \mc{1}{c}{V}&                 
 \mc{1}{c}{C}&                 \mc{1}{c}{C}&                 \mc{1}{c}{C}&                 \mc{1}{c}{C}&                 
 &                             &                             &                             &                             
 &                             &                             &                             &                             
 &                             &                             &                             \\                            
 &                             &                             \mc{1}{c}{N}&                 &                             
 \mc{1}{c}{C}&                 &                             &                             &                             
 &                             &                             \mc{1}{c}{T}&                 \mc{1}{c}{C}&                 
 \mc{1}{c}{C}&                 &                             \mc{1}{c}{T}&                 \mc{1}{c}{D}&                 
 \mc{1}{c}{P}&                 \mc{1}{c}{P}&                 \mc{1}{c}{P}&                 \mc{1}{c}{P}&                 
 &                             &                             &                             &                             
 &                             &                             &                             &                             
 &                             &                             &                             \\                            
&&&&& &&&&& &&&&& &&&&& &&&&& &&&&& &\\                                         
\hline                                                                          
&&&&& &&&&& &&&&& &&&&& &&&&& &&&&& &\\                                         
 1528&96-180 21:08&0&1& 58& 4& 1& 0& 9&15& 0&1&0&36&31&1&0&0&0&1&3&
4.01111&148.3& 39.8& 8.07664&118&214&-25&14.1& 1.9&$  2.1\cdot 10^{-14}$&  10.5\\
 1529&96-181 18:46&3&3& 49&19&23&11& 6& 8& 6&0&1&41& 0&1&0&0&0&1&3&
4.01505&148.3& 39.7& 8.08240&106&221&-15&23.8& 1.6&$  5.8\cdot 10^{-13}$&   6.0\\
 1530&96-194 15:47&3&3& 53&19&25& 6& 5& 5& 6&0&1&44& 0&1&0&0&0&1&3&
4.07231&148.7& 38.6& 8.16676&104&226&-14&43.7& 1.6&$  8.6\cdot 10^{-14}$&   6.0\\
 1531&96-197 08:46&0&2&164&14&22& 0& 8& 6& 6&0&0&40& 0&1&0&0&0&1&3&
4.08381&148.7& 38.4& 8.18386&258& 52&-12&21.4& 1.9&$  5.0\cdot 10^{-13}$&  10.5\\
 1532&96-221 07:41&0&1& 59& 7&11& 0& 7& 7& 7&0&0& 4&24&1&0&0&0&1&3&
4.18630&149.3& 36.4& 8.33840& 86&242&  1&43.7& 1.6&$  2.9\cdot 10^{-15}$&   6.0\\
&&&&& &&&&& &&&&& &&&&& &&&&& &&&&& &\\                      \hline                                                      
 1533&96-254 03:25&1&1& 14& 6&10& 2&12&15& 0&1&1&19&31&1&0&0&0&1&3&
4.31903&150.1& 33.9& 8.54431&336& 13& 50& 3.4& 1.6&$  8.0\cdot 10^{-12}$&   6.0\\
 1534&96-261 21:47&3&4&105&24&30&18& 7& 4& 5&0&1&47& 0&1&0&0&0&1&3&
4.34905&150.2& 33.3& 8.59178& 91&249& -2&31.4& 1.8&$  1.8\cdot 10^{-12}$&   9.5\\
 1535&96-281 02:43&0&1& 91& 1& 2& 0&15&15& 0&1&0& 2&23&1&0&0&0&1&3&
4.42111&150.6& 31.9& 8.70704& 52&274& 30&11.8&11.8&$  2.3\cdot 10^{-14}$&5858.3\\
 1536&96-289 05:04&0&1&189& 3& 1& 0&12&15&15&1&0&42&31&1&0&0&0&1&3&
4.45087&150.8& 31.3& 8.75516&184&150&-66& 5.0& 1.9&$  3.3\cdot 10^{-13}$&  10.5\\
 1537&96-292 14:17&0&1& 65& 3& 8& 0& 9& 8& 8&0&0& 0& 0&1&0&0&0&1&3&
4.46307&150.8& 31.1& 8.77499&  9&325& 57&22.7& 1.6&$  9.4\cdot 10^{-15}$&   6.0\\
&&&&& &&&&& &&&&& &&&&& &&&&& &&&&& &\\                      \hline                                                      
 1538&96-301 15:54&1&3&148&19&13& 4& 5&15& 0&1&1&44&31&1&0&0&0&1&3&
4.49564&151.0& 30.4& 8.82816&120&242&-28&34.1& 1.9&$  5.1\cdot 10^{-14}$&  10.5\\
 1539&96-305 08:50&1&2&181& 8& 7& 6&11&15&15&1&1& 6&31&1&0&0&0&1&3&
4.50840&151.1& 30.2& 8.84909&164&200&-62& 7.8& 1.9&$  4.9\cdot 10^{-13}$&  10.5\\
 1540&96-308 22:27&1&3&149&20&15&14& 5& 0& 5&0&1&43& 0&1&0&0&0&1&3&
4.52106&151.1& 29.9& 8.86992&117&245&-25&34.1& 1.9&$  8.9\cdot 10^{-14}$&  10.5\\
 1541&96-309 00:26&3&2&179&14&21& 2& 8& 8& 6&0&1&39& 0&1&0&0&0&1&3&
4.52149&151.1& 29.9& 8.87063&159&209&-59&21.4& 1.9&$  4.2\cdot 10^{-13}$&  10.5\\
 1542&96-315 19:01&3&3&156&19&23&10& 7& 7& 6&0&1&42& 0&1&0&0&0&1&3&
4.54480&151.2& 29.5& 8.90911&123&242&-31&21.0& 1.6&$  9.6\cdot 10^{-13}$&   6.0\\
&&&&& &&&&& &&&&& &&&&& &&&&& &&&&& &\\                      \hline                                                      
 1543&96-320 22:16&3&2&173&12&20& 8& 8& 5& 6&0&1&37& 0&1&0&0&0&1&3&
4.56226&151.3& 29.1& 8.93806&144&228&-48&21.4& 1.9&$  2.6\cdot 10^{-13}$&  10.5\\
 1544&96-331 07:53&3&2&182&10&14& 3& 8& 8& 7&0&1&35& 0&1&0&0&0&1&3&
4.59700&151.5& 28.4& 8.99598&152&219&-54&28.0& 1.6&$  3.9\cdot 10^{-14}$&   6.0\\
 1545&96-335 04:17&3&4&136&26&49&20& 6& 4& 5&0&1&46& 0&1&0&0&0&1&?&
4.60978&151.6& 28.2& 9.01737& 86&262&  1&21.4& 1.9&$  1.7\cdot 10^{-11}$&  10.5\\
 1546&96-359 13:34&0&1&151& 3& 1& 0& 8&15& 0&1&0& 0& 0&1&0&0&0&1&3&
4.68756&152.0& 26.6& 9.14883& 94&257& -5&21.4& 1.9&$  3.9\cdot 10^{-15}$&  10.5\\
 1547&97-010 02:02&1&3& 65&19&13& 5& 6&15&15&1&1&44&31&1&0&0&0&1&3&
4.73775&152.2& 25.5& 9.23469&322& 36& 40&21.4& 1.9&$  2.6\cdot 10^{-13}$&  10.5\\
&&&&& &&&&& &&&&& &&&&& &&&&& &&&&& &\\                      \hline                                                      
 1548&97-015 13:29&3&3&154&20&25& 2& 5& 6& 6&0&1&43& 0&1&0&0&0&1&3&
4.75405&152.3& 25.2& 9.26273& 83&260&  3&35.4& 1.6&$  2.8\cdot 10^{-13}$&   6.0\\
 1549&97-028 01:22&0&1&252& 3& 1& 0& 9&15& 0&1&0& 0& 0&1&0&0&0&1&3&
4.79027&152.5& 24.4& 9.32537&207&108&-54&14.1& 1.9&$  1.8\cdot 10^{-14}$&  10.5\\
 1550&97-043 22:37&3&3&195&21&27& 6& 6& 5& 5&0&1&46& 0&1&0&0&0&1&3&
4.83467&152.7& 23.4& 9.40269&101&245&-11&34.6& 1.6&$  5.0\cdot 10^{-13}$&   6.0\\
 1551&97-060 20:57&3&2&250&12&20&10& 9& 9& 7&0&1&38& 0&1&0&0&0&1&3&
4.87993&152.9& 22.4& 9.48209&133&223&-39&14.1& 1.9&$  1.2\cdot 10^{-12}$&  10.5\\
 1552&97-068 01:39&0&2& 18& 9& 4& 0& 6&15& 0&1&0&22&31&1&0&0&0&1&3&
4.89877&153.0& 22.0& 9.51530&147&210&-50&43.5& 1.9&$  1.2\cdot 10^{-15}$&  10.5\\
&&&&& &&&&& &&&&& &&&&& &&&&& &&&&& &\\                      \hline                                                      
 1553&97-080 22:16&1&2& 46&13& 8& 9&10&15& 0&1&1& 4&31&1&0&0&0&1&3&
4.93132&153.2& 21.2& 9.57292&167&179&-65&10.4& 1.9&$  6.0\cdot 10^{-13}$&  10.5\\
 1554&97-111 23:30&0&2&255&12& 8& 0& 9&15& 0&1&0&19&31&1&0&0&0&1&3&
5.00504&153.6& 19.5& 9.57292& 76&242& 11&14.1& 1.9&$  2.5\cdot 10^{-13}$&  10.5\\
 1555&97-120 11:13&3&4&249&27&49&13& 8& 6& 5&0&1&47& 0&1&0&0&0&1&3&
5.02412&153.7& 19.0& 9.73865& 65&246& 21&10.4& 1.9&$  1.9\cdot 10^{-10}$&  10.5\\
 1556&97-132 16:55&0&3& 48&19&14& 0& 8&15& 0&1&0&47&31&1&0&0&0&1&3&
5.05077&153.8& 18.3& 9.78659&138&219&-46&10.4& 1.9&$  2.9\cdot 10^{-12}$&  10.5\\
 1557&97-145 15:34&3&3& 19&19&22& 8& 6& 5& 6&0&1&41& 0&1&0&0&0&1&3&
5.07795&154.0& 17.6& 9.83562& 94&236& -4&34.6& 1.6&$  1.4\cdot 10^{-13}$&   6.0\\
&&&&& &&&&& &&&&& &&&&& &&&&& &&&&& &\\                      \hline                                                      
 1558&97-147 15:07&0&1&246& 5& 4& 0&10&15& 0&1&0&14&30&1&0&0&0&1&3&
5.08203&154.0& 17.5& 9.84300& 54&250& 33&10.4& 1.9&$  7.9\cdot 10^{-14}$&  10.5\\
 1559&97-148 22:16&3&3& 32&21&25& 8& 5& 5& 5&0&1&44& 0&1&0&0&0&1&3&
5.08457&154.0& 17.4& 9.84758&111&231&-20&43.7& 1.6&$  1.4\cdot 10^{-13}$&   6.0\\
 1560&97-150 14:33&3&3& 11&19&24&11& 5& 6& 5&0&1&42& 0&1&0&0&0&1&3&
5.08785&154.0& 17.3& 9.85352& 82&240&  6&35.4& 1.6&$  1.8\cdot 10^{-13}$&   6.0\\
 1561&97-155 09:59&0&1& 47& 7& 6& 0&11&15& 0&1&0&25&31&1&0&0&0&1&3&
5.09760&154.1& 17.0& 9.87115&132&225&-40& 7.8& 1.9&$  3.4\cdot 10^{-13}$&  10.5\\
 1562&97-166 19:46&0&1&249& 7& 5& 0& 9&10& 0&1&0&46&14&1&0&0&0&1&3&
5.11973&154.2& 16.4& 9.91123& 53&251& 34&14.1& 1.9&$  6.4\cdot 10^{-14}$&  10.5\\
&&&&& &&&&& &&&&& &&&&& &&&&& &&&&& &\\                      \hline                                                      
 1563&97-170 02:14&1&2& 79& 9& 3& 5& 9&15& 0&1&1&13&31&1&0&0&0&1&3&
5.12590&154.3& 16.2& 9.92241&173&182&-77&14.1& 1.9&$  6.6\cdot 10^{-14}$&  10.5\\
 1564&97-183 13:43&3&2& 11&14&21& 8& 7& 7& 6&0&1&40& 0&1&0&0&0&1&3&
5.15079&154.4& 15.5& 9.96754& 75&244& 14&34.1& 1.9&$  8.3\cdot 10^{-14}$&  10.5\\
 1565&97-223 08:01&0&2& 87&10& 5& 0& 9&15& 0&1&0&41&31&1&0&0&0&1&3&
5.21721&154.8& 13.4&10.08808&156&229&-64&14.1& 1.9&$  1.1\cdot 10^{-13}$&  10.5\\
 1566&97-228 12:46&3&3& 33&19&24&25& 5& 6& 6&0&1&43& 0&1&0&0&0&1&3&
5.22522&154.9& 13.2&10.10260& 69&252& 19&35.4& 1.6&$  1.8\cdot 10^{-13}$&   6.0\\
 1567&97-231 15:51&1&3& 64&19&22&13& 6& 6& 0&1&1&99&16&1&0&0&0&1&3&
5.22991&154.9& 13.0&10.11109&104&246&-14&28.0& 1.6&$  2.8\cdot 10^{-13}$&   6.0\\
&&&&& &&&&& &&&&& &&&&& &&&&& &&&&& &\\                      \hline                                                      
 1568&97-236 04:53&0&1& 48& 1& 1& 0&15& 0&11&0&0&10&23&1&0&0&0&1&3&
5.23655&155.0& 12.8&10.12310& 74&252& 15&11.8&11.8&$  1.9\cdot 10^{-14}$&5858.3\\
 1569&97-237 07:43&3&6&249&57&49&13& 8& 6& 5&0&1&47& 0&1&1&0&0&1&3&
5.23819&155.0& 12.7&10.12607&344& 22& 67&10.4& 1.9&$  1.6\cdot 10^{-09}$&  10.5\\
 1570&97-238 10:04&3&3& 78&19&23& 5& 7& 7& 6&0&1&41& 0&1&0&0&0&1&3&
5.23982&155.0& 12.6&10.12902&104&247&-13&21.0& 1.6&$  9.6\cdot 10^{-13}$&   6.0\\
 1571&97-247 23:29&0&1& 92& 3& 2& 0& 8&15& 0&1&0& 0& 0&1&0&0&0&1&3&
5.25328&155.1& 12.2&10.15334& 81&253&  8&21.4& 1.9&$  4.6\cdot 10^{-15}$&  10.5\\
 1572&97-253 11:13&1&1& 96& 3& 4& 2&10&15& 0&1&1& 7&31&1&0&0&0&1&3&
5.26081&155.2& 11.9&10.16692& 72&255& 17&10.4& 1.9&$  5.7\cdot 10^{-14}$&  10.5\\
&&&&& &&&&& &&&&& &&&&& &&&&& &&&&& &\\                      \hline                                                      
 1573&97-274 21:29&2&3&140&20&25&24& 5& 5& 5&0&1&44& 0&1&0&0&0&1&3&
5.28842&155.4& 10.8&10.21649&105&253&-15&43.7& 1.6&$  1.1\cdot 10^{-13}$&   6.0\\
 1574&97-295 21:15&3&2&154& 8&13& 2& 7& 8& 7&0&1&34& 0&1&0&0&0&1&3&
5.31259&155.6&  9.7&10.25945&116&255&-25&35.4& 1.6&$  1.1\cdot 10^{-14}$&   6.0\\
 1575&97-300 08:40&3&4&  1&28&49&28&12&10& 8&0&1&15& 0&1&0&0&0&1&3&
5.31728&155.6&  9.5&10.26772&260& 70&-10& 2.0& 1.9&$  5.1\cdot 10^{-08}$&  10.5\\
 1576&97-308 20:54&3&3&170&19&22& 4& 8& 6& 6&0&1&41& 0&1&0&0&0&1&3&
5.32604&155.7&  9.1&10.28311&136&254&-45&19.5& 1.6&$  1.1\cdot 10^{-12}$&   6.0\\
 1577&97-313 18:51&0&1&189& 2& 2& 0&15& 9& 0&1&0&99&27&1&0&0&0&1&3&
5.33099&155.8&  8.8&10.29175&161&246&-70&11.8&11.8&$  2.8\cdot 10^{-14}$&5858.3\\
&&&&& &&&&& &&&&& &&&&& &&&&& &&&&& &\\                      \hline                                                      
 1578&97-325 01:21&3&2&144&11&19& 6& 9& 7& 7&0&1&36& 0&1&0&0&0&1&3&
5.34155&155.9&  8.3&10.31004& 97&260& -6&14.1& 1.9&$  7.7\cdot 10^{-13}$&  10.5\\
 1579&97-329 04:12&2&4&162&29&49&17& 7& 9& 5&0&1&46& 0&1&0&0&0&1&3&
5.34523&155.9&  8.1&10.31636&121&257&-30&14.1& 1.9&$  1.2\cdot 10^{-10}$&  10.5\\
 1580&97-343 12:22&1&3&134&19&12& 8& 7&15& 0&1&1& 9&31&1&0&0&0&1&3&
5.35722&156.0&  7.3&10.33670& 80&263&  9&14.1& 1.9&$  1.0\cdot 10^{-12}$&  10.5\\
 1581&97-353 03:25&1&4&178&28& 8&17& 7&15& 4&1&1&99& 1&1&0&0&0&1&3&
5.36454&156.1&  6.9&10.34888&139&256&-48&14.1& 1.9&$  2.5\cdot 10^{-12}$&  10.5\\
 1582&97-355 05:19&0&1&114& 4& 3& 0& 7& 7&11&0&0&12&23&1&0&0&0&1&3&
5.36599&156.2&  6.8&10.35127& 51&268& 38&34.1& 1.9&$  1.2\cdot 10^{-15}$&  10.5\\
\end{tabular*} \end{sidewaystable}                                                     
\clearpage                                                                      

\thispagestyle{empty}
\begin{sidewaystable}                                                                   
\tabcolsep1.2mm
\tiny \hspace{-2cm}    
\begin{tabular*}{23cm} {%@{\extracolsep{\fill}}                                  
rrrrrrrrrrrrrrrcrrrrcrrrrrrrrrrrrrrrrrrrrr}                                     
 \mc{1}{c}{No.}&               \mc{1}{c}{IMP. DATE}&         \mc{1}{c}{C}&                 \mc{1}{c}{AR}&                
 \mc{1}{c}{S}&                 \mc{1}{c}{IA}&                \mc{1}{c}{EA}&                \mc{1}{c}{CA}&                
 \mc{1}{c}{IT}&                \mc{1}{c}{ET}&                \mc{1}{c}{E}&                 \mc{1}{c}{E}&                 
 \mc{1}{c}{I}&                 \mc{1}{c}{PA}&                \mc{1}{c}{P}&                 \mc{1}{c}{E}&                 
 \mc{1}{c}{I}&                 \mc{1}{c}{E}&                 \mc{1}{c}{C}&                 \mc{1}{c}{P}&                 
 \mc{1}{c}{HV}&                \mc{1}{c}{R}&                 \mc{1}{c}{LON}&               \mc{1}{c}{LAT}&               
 \mc{1}{c}{D$_{\rm Jup}$}&     \mc{1}{c}{ROT}&               \mc{1}{c}{S$_{\rm LON}$}&     \mc{1}{c}{S$_{\rm LAT}$}&     
 \mc{1}{c}{V}&                 \mc{1}{c}{VEF}&               \mc{1}{c}{M}&                 \mc{1}{c}{MEF}\\              
 &                             &                             \mc{1}{c}{L}&                 &                             
 \mc{1}{c}{E}&                 &                             &                             &                             
 &                             &                             \mc{1}{c}{I}&                 \mc{1}{c}{I}&                 
 \mc{1}{c}{I}&                 &                             \mc{1}{c}{E}&                 \mc{1}{c}{V}&                 
 \mc{1}{c}{C}&                 \mc{1}{c}{C}&                 \mc{1}{c}{C}&                 \mc{1}{c}{C}&                 
 &                             &                             &                             &                             
 &                             &                             &                             &                             
 &                             &                             &                             \\                            
 &                             &                             \mc{1}{c}{N}&                 &                             
 \mc{1}{c}{C}&                 &                             &                             &                             
 &                             &                             \mc{1}{c}{T}&                 \mc{1}{c}{C}&                 
 \mc{1}{c}{C}&                 &                             \mc{1}{c}{T}&                 \mc{1}{c}{D}&                 
 \mc{1}{c}{P}&                 \mc{1}{c}{P}&                 \mc{1}{c}{P}&                 \mc{1}{c}{P}&                 
 &                             &                             &                             &                             
 &                             &                             &                             &                             
 &                             &                             &                             \\                            
&&&&& &&&&& &&&&& &&&&& &&&&& &&&&& &\\                                         
\hline                                                                          
&&&&& &&&&& &&&&& &&&&& &&&&& &&&&& &\\                                         
 1583&97-358 10:24&0&1&188& 2& 3& 0&11&11& 6&0&0& 4&13&1&0&0&0&1&3&
5.36829&156.2&  6.6&10.35503&154&253&-63&11.8&11.8&$  3.2\cdot 10^{-14}$&5858.3\\
 1584&97-358 14:00&0&1&206& 6&10& 0& 8& 8& 9&0&0& 0& 0&1&0&0&0&1&3&
5.36838&156.2&  6.6&10.35518&179&190&-87&28.0& 1.6&$  1.0\cdot 10^{-14}$&   6.0\\
 1585&97-359 02:25&3&4& 68&29&49&20& 9& 7& 5&0&1& 1& 0&1&0&0&0&1&3&
5.36872&156.2&  6.6&10.35574&345& 38& 71& 7.8& 1.9&$  5.4\cdot 10^{-10}$&  10.5\\
 1586&97-361 19:23&3&2&141& 9&14& 4& 9& 9& 9&0&1&36& 0&1&0&0&0&1&3&
5.37061&156.2&  6.4&10.35881& 87&262&  3&21.0& 1.6&$  9.8\cdot 10^{-14}$&   6.0\\
 1587&98-008 01:41&3&2&164&11&20& 7& 8& 6& 6&0&1&37& 0&1&0&0&0&1&3&
5.37784&156.3&  5.9&10.37035&118&257&-27&21.4& 1.9&$  2.2\cdot 10^{-13}$&  10.5\\
&&&&& &&&&& &&&&& &&&&& &&&&& &&&&& &\\                      \hline                                                      
 1588&98-022 16:15&3&2&175&11&20& 3& 7& 5& 7&0&1&37& 0&1&0&0&0&1&3&
5.38608&156.5&  5.2&10.38297&130&255&-39&34.1& 1.9&$  4.4\cdot 10^{-14}$&  10.5\\
 1589&98-025 12:28&0&1&118& 1& 4& 0&15& 4& 0&1&0& 0& 0&1&0&0&0&1&3&
5.38755&156.5&  5.0&10.38513& 49&264& 39&11.8&11.8&$  3.0\cdot 10^{-14}$&5858.3\\
 1590&98-029 12:06&3&3&129&19&24& 8& 5& 6& 6&0&1&43& 0&1&0&0&0&1&3&
5.38950&156.5&  4.8&10.38797& 64&261& 25&35.4& 1.6&$  1.8\cdot 10^{-13}$&   6.0\\
 1591&98-036 00:05&0&1&211& 3& 4& 0&11&14& 0&1&0& 3&23&1&0&0&0&1&3&
5.39247&156.6&  4.5&10.39215&176&245&-85& 7.8& 1.9&$  1.2\cdot 10^{-13}$&  10.5\\
 1592&98-039 10:16&1&3&129&20&21& 4& 4&14& 0&1&1&46&31&1&0&0&0&1&3&
5.39391&156.6&  4.4&10.39410& 59&259& 30&43.5& 1.9&$  6.3\cdot 10^{-14}$&  10.5\\
&&&&& &&&&& &&&&& &&&&& &&&&& &&&&& &\\                      \hline                                                      
 1593&98-069 04:13&0&1&189& 1& 3& 0&15&13&11&0&0&11&23&1&0&0&0&1&3&
5.40358&156.9&  2.9&10.40511&350& 25& 77&11.8&11.8&$  2.6\cdot 10^{-14}$&5858.3\\
 1594&98-080 22:21&3&3& 20&21&26&28& 5& 5& 5&0&1&45& 0&1&0&0&0&1&3&
5.40591&157.0&  2.3&10.40636&106&247&-15&43.7& 1.6&$  1.6\cdot 10^{-13}$&   6.0\\
 1595&98-083 23:56&3&3& 20&23&29& 4& 3& 3& 5&0&1&46& 0&1&0&0&0&1&3&
5.40638&157.0&  2.2&10.40640&104&246&-13&70.0& 1.6&$  5.5\cdot 10^{-14}$&   6.0\\
 1596&98-088 22:15&0&1& 50& 7&11& 0& 9& 9& 8&0&0& 0& 0&1&0&0&0&1&3&
5.40702&157.1&  2.0&10.40620&145&246&-54&21.0& 1.6&$  4.4\cdot 10^{-14}$&   6.0\\
 1597&98-091 16:54&0&1& 86& 2& 6& 0& 9& 8& 9&0&0& 6&31&1&0&0&0&1&3&
5.40731&157.1&  1.8&10.40596&194& 48&-74&36.7& 2.0&$  1.1\cdot 10^{-15}$&  12.5\\
&&&&& &&&&& &&&&& &&&&& &&&&& &&&&& &\\                      \hline                                                      
 1598&98-094 03:55&1&2& 46&12&12& 8& 8& 8& 0&1&1& 0& 0&1&0&0&0&1&3&
5.40753&157.1&  1.7&10.40566&137&245&-46&21.4& 1.9&$  1.1\cdot 10^{-13}$&  10.5\\
 1599&98-095 06:47&0&1&248& 3& 4& 0&10&15& 0&1&0&39&31&1&0&0&0&1&3&
5.40762&157.1&  1.6&10.40550& 61&246& 28&10.4& 1.9&$  5.7\cdot 10^{-14}$&  10.5\\
 1600&98-097 22:11&3&3& 27&20&24&17& 5& 6& 6&0&1&42& 0&1&0&0&0&1&3&
5.40780&157.2&  1.5&10.40506&112&244&-21&35.4& 1.6&$  2.4\cdot 10^{-13}$&   6.0\\
 1601&98-099 06:11&1&1&241& 4& 5& 2&12&15& 0&1&1&11&31&1&0&0&0&1&3&
5.40787&157.2&  1.5&10.40479& 50&247& 40& 5.0& 1.9&$  7.0\cdot 10^{-13}$&  10.5\\
 1602&98-102 15:22&1&1& 70& 1& 2& 1&15&15& 0&1&1& 3&20&1&0&0&0&1&3&
5.40801&157.2&  1.3&10.40403&171&260&-79&11.8&11.8&$  2.3\cdot 10^{-14}$&5858.3\\
&&&&& &&&&& &&&&& &&&&& &&&&& &&&&& &\\                      \hline                                                      
 1603&98-139 06:14&1&1& 37& 3& 5& 5&14&13& 0&1&1& 0& 0&1&0&0&0&1&3&
5.40503&157.6& -0.5&10.38651&122&243&-31& 3.9& 1.6&$  1.3\cdot 10^{-12}$&   6.0\\
 1604&98-139 14:44&1&1&  3& 4& 3& 4&10&15& 0&1&1& 9&31&1&0&0&0&1&3&
5.40498&157.6& -0.5&10.38633& 74&242& 16&10.4& 1.9&$  5.8\cdot 10^{-14}$&  10.5\\
 1605&98-139 17:34&1&3&  5&19&20& 3& 8& 7& 0&1&1&21& 4&1&0&0&0&1&3&
5.40495&157.6& -0.5&10.38624& 77&242& 13&10.4& 1.9&$  5.3\cdot 10^{-12}$&  10.5\\
 1606&98-145 13:37&0&1&159& 6&11& 0& 7& 7& 8&0&0& 0& 0&1&0&0&0&1&3&
5.40370&157.6& -0.8&10.38182&295& 51& 25&43.7& 1.6&$  2.3\cdot 10^{-15}$&   6.0\\
 1607&98-147 20:24&1&1&249& 3& 4& 2&11&13& 0&1&1& 9&31&1&0&0&0&1&3&
5.40316&157.6& -0.9&10.38000& 60&242& 30& 7.8& 1.9&$  1.2\cdot 10^{-13}$&  10.5\\
&&&&& &&&&& &&&&& &&&&& &&&&& &&&&& &\\                      \hline                                                      
 1608&98-153 11:54&1&1& 18& 7& 5& 5&12&15& 0&1&1&11&31&1&0&0&0&1&3&
5.40169&157.7& -1.2&10.37519& 97&242& -6& 5.0& 1.9&$  1.2\cdot 10^{-12}$&  10.5\\
 1609&98-161 07:21&3&3& 18&19&24&11& 6& 6& 5&0&1&42& 0&1&0&0&0&1&3&
5.39930&157.8& -1.6&10.36778& 97&243& -6&28.0& 1.6&$  3.8\cdot 10^{-13}$&   6.0\\
 1610&98-162 23:44&0&1& 10& 6&19& 0&10& 4& 8&0&0&39& 0&1&0&0&0&1&3&
5.39877&157.8& -1.6&10.36615& 85&243&  5&10.4& 1.9&$  7.0\cdot 10^{-13}$&  10.5\\
 1611&98-173 13:59&3&1& 12& 7&11& 4& 7& 7& 6&0&1& 0& 0&1&0&0&0&1&3&
5.39484&157.9& -2.2&10.35469& 89&244&  1&43.7& 1.6&$  2.9\cdot 10^{-15}$&   6.0\\
 1612&98-182 19:15&0&1&152& 4& 1& 0& 8&15& 0&1&0& 5&23&1&0&0&0&1&3&
5.39086&158.0& -2.6&10.34354&287& 55& 17&21.4& 1.9&$  4.5\cdot 10^{-15}$&  10.5\\
&&&&& &&&&& &&&&& &&&&& &&&&& &&&&& &\\                      \hline                                                      
 1613&98-200 16:10&0&2&132& 9&15& 0& 7& 8& 6&0&0&36& 0&1&0&0&0&1&3&
5.38168&158.1& -3.5&10.31893&261& 56& -8&34.1& 1.9&$  2.0\cdot 10^{-14}$&  10.5\\
 1614&98-201 23:14&0&1& 21& 4&19& 0&10& 6&13&0&0&36& 0&1&0&0&0&1&3&
5.38097&158.1& -3.5&10.31705&105&248&-14&10.4& 1.9&$  5.0\cdot 10^{-13}$&  10.5\\
 1615&98-202 13:44&1&2&231&11&13& 6& 9&15& 0&1&1&24&31&1&0&0&0&1&3&
5.38061&158.1& -3.6&10.31611& 41&246& 49& 5.7& 2.1&$  5.8\cdot 10^{-12}$&  15.7\\
 1616&98-205 22:27&3&3&242&19&23& 3& 6& 7& 6&0&1&41& 0&1&0&0&0&1&3&
5.37862&158.2& -3.7&10.31093& 57&246& 33&25.9& 1.6&$  4.3\cdot 10^{-13}$&   6.0\\
 1617&98-207 01:42&0&2&252&10&14& 0& 7& 8& 6&0&0&35& 0&1&0&0&0&1&3&
5.37794&158.2& -3.8&10.30917& 71&247& 20&35.4& 1.6&$  1.8\cdot 10^{-14}$&   6.0\\
&&&&& &&&&& &&&&& &&&&& &&&&& &&&&& &\\                      \hline                                                      
 1618&98-210 03:24&0&1& 46& 6&10& 0& 8& 8& 9&0&0& 4&31&1&0&0&0&1&3&
5.37601&158.2& -3.9&10.30420&143&255&-51&28.0& 1.6&$  1.0\cdot 10^{-14}$&   6.0\\
 1619&98-214 02:36&3&3&253&19&22&14&10&11& 7&0&1&37& 0&1&0&0&0&1&3&
5.37354&158.3& -4.1&10.29787& 75&248& 16& 6.2& 1.6&$  3.2\cdot 10^{-11}$&   6.0\\
 1620&98-220 18:13&3&3&  8&20&25& 4& 5& 6& 5&0&1&44& 0&1&0&0&0&1&3&
5.36901&158.3& -4.5&10.28639& 93&250& -2&35.4& 1.6&$  2.8\cdot 10^{-13}$&   6.0\\
 1621&98-222 10:05&3&3&251&19&25& 6& 6& 7& 5&0&1&43& 0&1&0&0&0&1&3&
5.36788&158.3& -4.5&10.28354& 76&249& 14&25.9& 1.6&$  5.9\cdot 10^{-13}$&   6.0\\
 1622&98-224 20:40&0&1& 20& 2& 9& 0&15&15& 0&1&0&99&31&1&0&0&0&1&3&
5.36619&158.4& -4.7&10.27932&111&252&-19&11.8&11.8&$  8.8\cdot 10^{-14}$&5858.3\\
&&&&& &&&&& &&&&& &&&&& &&&&& &&&&& &\\                      \hline                                                      
 1623&98-232 21:14&3&3& 36&20&25& 1& 5& 5& 6&0&1&45& 0&1&0&0&0&1&3&
5.36017&158.4& -5.1&10.26433&146&262&-54&43.7& 1.6&$  1.1\cdot 10^{-13}$&   6.0\\
 1624&98-233 11:26&0&1&239& 7& 8& 0&13&10& 0&1&0&99&22&1&0&0&0&1&3&
5.35979&158.4& -5.1&10.26338& 73&251& 18& 2.5& 1.9&$  2.1\cdot 10^{-11}$&  10.5\\
 1625&98-233 18:56&0&1&220& 2& 1& 0&11&15&14&0&0&99& 4&1&0&0&0&1&3&
5.35950&158.4& -5.1&10.26267& 46&250& 44&11.8&11.8&$  2.4\cdot 10^{-14}$&5858.3\\
 1626&98-236 02:46&1&2& 15& 8&12& 8&12&15& 0&1&1&20&31&1&0&0&0&1&3&
5.35775&158.5& -5.2&10.25834&124&256&-32& 3.4& 1.6&$  1.6\cdot 10^{-11}$&   6.0\\
 1627&98-239 17:56&3&2&229&14&22&13& 9& 9& 7&0&1&40& 0&1&0&0&0&1&3&
5.35486&158.5& -5.4&10.25125& 90&253&  1&14.1& 1.9&$  2.3\cdot 10^{-12}$&  10.5\\
&&&&& &&&&& &&&&& &&&&& &&&&& &&&&& &\\                      \hline                                                      
 1628&98-241 07:52&3&2&244& 9&14& 2& 9& 9& 8&0&1& 0& 0&1&0&0&0&1&3&
5.35354&158.5& -5.5&10.24802&126&257&-34&21.0& 1.6&$  9.8\cdot 10^{-14}$&   6.0\\
 1629&98-249 22:07&1&2&200&13&10&10& 8& 9& 0&1&1& 0& 0&1&0&0&0&1&3&
5.34625&158.6& -5.9&10.23030&121&258&-29&21.4& 1.9&$  9.0\cdot 10^{-14}$&  10.5\\
 1630&98-251 04:52&1&2&180&12& 5& 6& 8&15& 0&1&1& 0& 0&1&0&0&0&1&3&
5.34516&158.6& -6.0&10.22765& 97&255& -6&21.4& 1.9&$  3.1\cdot 10^{-14}$&  10.5\\
 1631&98-254 22:37&3&3&151&21&26&10& 5& 6& 5&0&1&44& 0&1&0&0&0&1&3&
5.34182&158.6& -6.2&10.21960& 68&253& 23&35.4& 1.6&$  3.9\cdot 10^{-13}$&   6.0\\
 1632&98-262 10:12&3&1&160& 7&11& 1& 8& 8& 7&0&1& 2&23&1&0&0&0&1&3&
5.33488&158.7& -6.5&10.20298& 96&257& -4&28.0& 1.6&$  1.5\cdot 10^{-14}$&   6.0\\
&&&&& &&&&& &&&&& &&&&& &&&&& &&&&& &\\                      \hline                                                      
 1633&98-273 09:10&3&2&143&14&21& 8& 8& 7& 6&0&1&40& 0&1&0&0&0&1&3&
5.32406&158.8& -7.1&10.17733& 77&257& 14&21.4& 1.9&$  4.2\cdot 10^{-13}$&  10.5\\
 1634&98-283 05:38&1&3&229&18&11&13&12&15& 0&1&1&99&31&1&0&0&0&1&3&
5.31385&158.9& -7.6&10.15335&203& 49&-63& 2.0& 1.9&$  3.3\cdot 10^{-10}$&  10.5\\
 1635&98-287 03:20&3&3&155&20&24&20& 5& 5& 5&0&1&43& 0&1&0&0&0&1&3&
5.30949&159.0& -7.8&10.14317&100&262& -8&43.7& 1.6&$  9.8\cdot 10^{-14}$&   6.0\\
 1636&98-311 16:27&3&3&148&19&22&10& 9& 9& 6&0&1&39& 0&1&0&0&0&1&3&
5.28059&159.2& -9.0&10.07655& 93&263& -1&12.6& 1.6&$  4.7\cdot 10^{-12}$&   6.0\\
 1637&98-319 10:27&3&2&125&13&20& 5& 9& 7& 8&0&1&38& 0&1&0&0&0&1&3&
5.27067&159.3& -9.4&10.05396& 62&260& 28&14.1& 1.9&$  1.4\cdot 10^{-12}$&  10.5\\
\end{tabular*} \end{sidewaystable}                                                     
\clearpage                                                                      
\thispagestyle{empty}
\begin{sidewaystable}                                                                   
\tabcolsep1.2mm
\tiny \hspace{-2cm}        
\begin{tabular*}{23cm} {%@{\extracolsep{\fill}}                                  
rrrrrrrrrrrrrrrcrrrrcrrrrrrrrrrrrrrrrrrrrr}                                     
 \mc{1}{c}{No.}&               \mc{1}{c}{IMP. DATE}&         \mc{1}{c}{C}&                 \mc{1}{c}{AR}&                
 \mc{1}{c}{S}&                 \mc{1}{c}{IA}&                \mc{1}{c}{EA}&                \mc{1}{c}{CA}&                
 \mc{1}{c}{IT}&                \mc{1}{c}{ET}&                \mc{1}{c}{E}&                 \mc{1}{c}{E}&                 
 \mc{1}{c}{I}&                 \mc{1}{c}{PA}&                \mc{1}{c}{P}&                 \mc{1}{c}{E}&                 
 \mc{1}{c}{I}&                 \mc{1}{c}{E}&                 \mc{1}{c}{C}&                 \mc{1}{c}{P}&                 
 \mc{1}{c}{HV}&                \mc{1}{c}{R}&                 \mc{1}{c}{LON}&               \mc{1}{c}{LAT}&               
 \mc{1}{c}{D$_{\rm Jup}$}&     \mc{1}{c}{ROT}&               \mc{1}{c}{S$_{\rm LON}$}&     \mc{1}{c}{S$_{\rm LAT}$}&     
 \mc{1}{c}{V}&                 \mc{1}{c}{VEF}&               \mc{1}{c}{M}&                 \mc{1}{c}{MEF}\\              
 &                             &                             \mc{1}{c}{L}&                 &                             
 \mc{1}{c}{E}&                 &                             &                             &                             
 &                             &                             \mc{1}{c}{I}&                 \mc{1}{c}{I}&                 
 \mc{1}{c}{I}&                 &                             \mc{1}{c}{E}&                 \mc{1}{c}{V}&                 
 \mc{1}{c}{C}&                 \mc{1}{c}{C}&                 \mc{1}{c}{C}&                 \mc{1}{c}{C}&                 
 &                             &                             &                             &                             
 &                             &                             &                             &                             
 &                             &                             &                             \\                            
 &                             &                             \mc{1}{c}{N}&                 &                             
 \mc{1}{c}{C}&                 &                             &                             &                             
 &                             &                             \mc{1}{c}{T}&                 \mc{1}{c}{C}&                 
 \mc{1}{c}{C}&                 &                             \mc{1}{c}{T}&                 \mc{1}{c}{D}&                 
 \mc{1}{c}{P}&                 \mc{1}{c}{P}&                 \mc{1}{c}{P}&                 \mc{1}{c}{P}&                 
 &                             &                             &                             &                             
 &                             &                             &                             &                             
 &                             &                             &                             \\                            
&&&&& &&&&& &&&&& &&&&& &&&&& &&&&& &\\                                         
\hline                                                                          
&&&&& &&&&& &&&&& &&&&& &&&&& &&&&& &\\                                         
 1638&98-350 21:24&0&1&143& 1& 8& 0&15& 7&12&0&0& 0& 0&1&0&0&0&1&3&
5.22641&159.6&-11.0& 9.95466& 91&266&  0&11.8&11.8&$  6.1\cdot 10^{-14}$&5858.3\\
 1639&98-363 17:29&1&5&163&55&56&30&14&12& 0&1&1&33&31&1&0&0&0&1&3&
5.20671&159.7&-11.7& 9.91111&125&275&-31& 2.0& 1.9&$  8.2\cdot 10^{-07}$&  10.5\\
 1640&99-007 22:18&0&2&168& 8& 7& 0& 9&15& 0&1&0&10&31&1&0&0&0&1&3&
5.19176&159.8&-12.2& 9.87831&135&279&-40&14.1& 1.9&$  1.1\cdot 10^{-13}$&  10.5\\
 1641&99-008 15:37&0&1& 85& 5& 9& 0&11&15& 0&1&0&19&31&1&0&0&0&1&3&
5.19053&159.8&-12.2& 9.87561& 18&235& 71& 4.2& 1.6&$  2.8\cdot 10^{-12}$&   6.0\\
 1642&99-013 00:20&0&1&147& 2& 3& 0&10&15& 0&1&0& 1&31&1&0&0&0&1&3&
5.18324&159.9&-12.4& 9.85971&106&268&-13&11.8&11.8&$  3.2\cdot 10^{-14}$&5858.3\\
&&&&& &&&&& &&&&& &&&&& &&&&& &&&&& &\\                      \hline                                                      
 1643&99-013 06:04&3&2&129&15&22& 2& 8& 7& 6&0&1&41& 0&1&0&0&0&1&3&
5.18282&159.9&-12.5& 9.85879& 81&262& 10&21.4& 1.9&$  5.4\cdot 10^{-13}$&  10.5\\
 1644&99-016 08:38&1&2& 48&12&10& 6& 9&15& 0&1&1&40&31&1&0&0&0&1&3&
5.17775&159.9&-12.6& 9.84775&329& 93& 58&14.1& 1.9&$  3.5\cdot 10^{-13}$&  10.5\\
 1645&99-040 14:08&0&1&114& 3& 7& 0& 8& 8& 9&0&0& 0& 0&1&0&0&0&1&3&
5.13463&160.2&-13.9& 9.75469& 82&258&  9&28.0& 1.6&$  3.7\cdot 10^{-15}$&   6.0\\
 1646&99-052 07:32&1&1& 63& 5& 2& 2& 9&15& 0&1&1&12&31&1&0&0&0&1&3&
5.11236&160.3&-14.5& 9.70717& 36&237& 52&14.1& 1.9&$  2.9\cdot 10^{-14}$&  10.5\\
 1647&99-058 06:44&0&1&246& 2& 1& 0&15&15& 0&1&0&36&31&1&0&0&0&1&3&
5.10064&160.4&-14.9& 9.68230&314& 82& 43&11.8&11.8&$  2.4\cdot 10^{-14}$&5858.3\\
&&&&& &&&&& &&&&& &&&&& &&&&& &&&&& &\\                      \hline                                                      
 1648&99-059 11:10&2&3& 94&21&26&13& 9& 8& 6&0&1&45& 0&1&0&0&0&1&3&
5.09842&160.4&-14.9& 9.67760&108&261&-14&14.4& 1.6&$  9.9\cdot 10^{-12}$&   6.0\\
 1649&99-080 14:23&3&3& 31&23&29&26& 7& 5& 5&0&1&47& 0&1&0&0&0&1&3&
5.05510&160.6&-16.1& 9.58659& 79&247& 12&28.1& 1.6&$  1.9\cdot 10^{-12}$&   6.0\\
 1650&99-081 21:16&1&2& 41&11& 8& 9& 9&15& 0&1&1& 4&23&1&0&0&0&1&3&
5.05218&160.6&-16.2& 9.58049& 94&252& -1&14.1& 1.9&$  2.1\cdot 10^{-13}$&  10.5\\
 1651&99-082 04:35&0&1&218& 1& 4& 0&15& 0&13&0&0& 0& 0&1&0&0&0&1&3&
5.05165&160.6&-16.2& 9.57938&343&105& 68&11.8&11.8&$  3.0\cdot 10^{-14}$&5858.3\\
 1652&99-088 03:51&0&1& 73& 6&11& 0& 7& 8& 6&0&0& 0& 0&1&0&0&0&1&3&
5.03874&160.7&-16.5& 9.55253&146&278&-48&35.4& 1.6&$  5.7\cdot 10^{-15}$&   6.0\\
&&&&& &&&&& &&&&& &&&&& &&&&& &&&&& &\\                      \hline                                                      
 1653&99-088 15:36&3&3& 93&21&26& 3& 5& 5& 5&0&1&46& 0&1&0&0&0&1&3&
5.03765&160.7&-16.5& 9.55027&175&322&-64&43.7& 1.6&$  1.6\cdot 10^{-13}$&   6.0\\
 1654&99-094 09:48&0&1& 63& 6&10& 0& 7& 7& 7&0&0& 0& 0&1&0&0&0&1&3&
5.02504&160.8&-16.9& 9.52414&138&270&-41&43.7& 1.6&$  2.0\cdot 10^{-15}$&   6.0\\
 1655&99-104 17:43&0&1&245& 2& 2& 0&15&15& 0&1&0& 8&31&1&0&0&0&1&3&
5.00200&160.9&-17.4& 9.47667& 41&226& 48&11.8&11.8&$  2.8\cdot 10^{-14}$&5858.3\\
 1656&99-112 20:59&3&2& 22&12&20& 2& 8& 6& 7&0&1&37& 0&1&0&0&0&1&3&
4.98323&161.0&-17.9& 9.43825& 91&245&  1&21.4& 1.9&$  2.6\cdot 10^{-13}$&  10.5\\
 1657&99-115 05:54&0&2&197& 8& 1& 0& 5&15&11&0&0& 4&22&1&0&0&0&1&3&
4.97766&161.0&-18.0& 9.42689&338& 93& 64&70.0& 1.9&$  9.6\cdot 10^{-17}$&  10.5\\
&&&&& &&&&& &&&&& &&&&& &&&&& &&&&& &\\                      \hline                                                      
 1658&99-137 16:00&3&1&156& 7&12& 5& 7& 7& 6&0&1& 0& 0&1&0&0&0&1&3&
4.92299&161.3&-19.3& 9.31633&289& 61& 20&43.7& 1.6&$  3.3\cdot 10^{-15}$&   6.0\\
 1659&99-150 20:38&0&2& 10& 9&19& 0& 9& 8& 9&0&0&37& 0&1&0&0&0&1&3&
4.88949&161.5&-20.1& 9.24944& 88&243&  4&14.1& 1.9&$  5.7\cdot 10^{-13}$&  10.5\\
 1660&99-152 12:17&0&1& 94& 5& 6& 0& 8&15& 0&1&0&39&31&1&0&0&0&1&3&
4.88494&161.5&-20.2& 9.24039&206& 18&-53&21.4& 1.9&$  1.2\cdot 10^{-14}$&  10.5\\
 1661&99-157 13:23&0&2& 20&10&14& 0& 8& 8& 6&0&0&35& 0&1&0&0&0&1&3&
4.87180&161.6&-20.5& 9.21437&105&249&-11&28.0& 1.6&$  3.9\cdot 10^{-14}$&   6.0\\
 1662&99-165 01:16&3&2& 32&13&21&14& 7& 5& 6&0&1&39& 0&1&0&0&0&1&3&
4.85177&161.7&-21.0& 9.17487&124&258&-29&34.1& 1.9&$  7.1\cdot 10^{-14}$&  10.5\\
&&&&& &&&&& &&&&& &&&&& &&&&& &&&&& &\\                      \hline                                                      
 1663&99-179 00:37&0&2& 94& 8& 8& 0&14&15& 0&1&0&41&27&1&0&0&0&1&3&
4.81332&161.9&-21.8& 9.09970&216& 30&-46& 2.0& 1.9&$  6.4\cdot 10^{-11}$&  10.5\\
 1664&99-181 00:58&0&2& 74& 8&13& 0& 8& 8& 7&0&0&35& 0&1&0&0&0&1&3&
4.80772&161.9&-21.9& 9.08880&189&351&-63&28.0& 1.6&$  2.4\cdot 10^{-14}$&   6.0\\
 1665&99-196 11:11&1&2&255&14&22& 4&13&12& 0&1&1&31&31&1&0&0&0&1&3&
4.76368&162.1&-22.9& 9.00382& 91&249&  1& 2.5& 1.9&$  4.3\cdot 10^{-10}$&  10.5\\
 1666&99-198 18:22&0&1& 17& 3& 8& 0& 8& 8& 9&0&0& 0& 0&1&0&0&0&1&3&
4.75673&162.1&-23.0& 8.99049&117&259&-22&28.0& 1.6&$  4.5\cdot 10^{-15}$&   6.0\\
 1667&99-218 18:45&3&3&246&19&23& 8& 9& 9& 6&0&1&40& 0&1&0&0&0&1&3&
4.69656&162.4&-24.3& 8.87630& 95&253& -2&12.6& 1.6&$  5.4\cdot 10^{-12}$&   6.0\\
&&&&& &&&&& &&&&& &&&&& &&&&& &&&&& &\\                      \hline                                                      
 1668&99-222 09:16&0&1&199& 5&10& 0& 7& 7& 8&0&0&10&23&1&0&0&0&1&3&
4.68535&162.5&-24.5& 8.85522& 33&226& 54&43.7& 1.6&$  1.7\cdot 10^{-15}$&   6.0\\
 1669&99-223 00:23&0&1&238& 4& 2& 0& 9&15& 0&1&0&34&31&1&0&0&0&1&3&
4.68341&162.5&-24.6& 8.85158& 88&252&  4&14.1& 1.9&$  2.5\cdot 10^{-14}$&  10.5\\
 1670&99-225 21:14&1&3&207&20& 8&11& 7&15& 0&1&1&45&31&1&0&0&0&1&?&
4.67443&162.5&-24.7& 8.83477& 49&237& 40&14.1& 1.9&$  7.0\cdot 10^{-13}$&  10.5\\
 1671&99-227 07:11&1&2&249&12&22& 9&10&14& 0&1&1&25&31&1&0&0&0&1&3&
4.67012&162.6&-24.8& 8.82670&111&261&-17&10.4& 1.9&$  3.4\cdot 10^{-12}$&  10.5\\
 1672&99-235 16:00&3&3& 21&23&28& 5& 6& 4& 5&0&1&46& 0&1&0&0&0&1&3&
4.64354&162.7&-25.4& 8.77724&161&302&-57&38.7& 1.6&$  5.2\cdot 10^{-13}$&   6.0\\
&&&&& &&&&& &&&&& &&&&& &&&&& &&&&& &\\                      \hline                                                      
 1673&99-242 11:53&0&1&207& 2& 1& 0&15&15& 0&1&0& 0& 0&1&0&0&0&1&3&
4.62174&162.8&-25.8& 8.73695& 80&253& 11&11.8&11.8&$  2.4\cdot 10^{-14}$&5858.3\\
 1674&99-257 06:03&0&1&231& 2& 2& 0&15&15& 0&1&0& 4&23&1&0&0&0&1&3&
4.57252&163.0&-26.8& 8.64687&144&288&-45&11.8&11.8&$  2.8\cdot 10^{-14}$&5858.3\\
 1675&99-270 12:05&3&1&157& 4& 9&13& 7& 7& 7&0&1& 0& 0&1&0&0&0&1&3&
4.52727&163.2&-27.7& 8.56519& 58&249& 31&43.7& 1.6&$  1.2\cdot 10^{-15}$&   6.0\\
 1676&99-272 03:19&0&2&159&14&21& 0& 8& 5& 7&0&0&43& 0&1&0&0&0&1&3&
4.52163&163.3&-27.8& 8.55508& 65&252& 25&21.4& 1.9&$  4.2\cdot 10^{-13}$&  10.5\\
 1677&99-272 21:01&3&4&220&27&49&11& 8& 6& 5&0&1&46& 0&1&0&0&0&1&3&
4.51902&163.3&-27.9& 8.55041&150&298&-49&10.4& 1.9&$  1.9\cdot 10^{-10}$&  10.5\\
&&&&& &&&&& &&&&& &&&&& &&&&& &&&&& &\\                      \hline                                                      
 1678&99-273 11:55&3&2& 13&10&14&17& 7& 8& 6&0&1&36& 0&1&0&0&0&1&3&
4.51727&163.3&-27.9& 8.54730&219& 46&-42&35.4& 1.6&$  1.8\cdot 10^{-14}$&   6.0\\
 1679&99-278 10:02&0&2&115&10& 2& 0& 9&15& 0&1&0&37&31&1&0&0&0&1&3&
4.49975&163.4&-28.3& 8.51603& 10&197& 69&14.1& 1.9&$  6.8\cdot 10^{-14}$&  10.5\\
 1680&99-280 03:07&3&3&125&23&30& 8& 8& 4& 5&0&1&46& 0&1&0&0&0&1&3&
4.49357&163.4&-28.4& 8.50504& 23&222& 61&10.4& 1.9&$  5.8\cdot 10^{-11}$&  10.5\\
 1681&99-282 09:17&0&1&198& 5& 9& 0& 8& 8& 8&0&0& 9&31&1&0&0&0&1&3&
4.48559&163.5&-28.5& 8.49089&129&283&-31&28.0& 1.6&$  7.4\cdot 10^{-15}$&   6.0\\
 1682&99-286 02:42&0&1&239& 6&10& 0& 8& 8& 7&0&0& 0& 0&1&0&0&0&1&3&
4.47265&163.5&-28.8& 8.46800&189&  7&-59&28.0& 1.6&$  1.0\cdot 10^{-14}$&   6.0\\
&&&&& &&&&& &&&&& &&&&& &&&&& &&&&& &\\                      \hline                                                      
 1683&99-304 03:04&0&1& 43& 7&13& 0& 7& 7& 7&0&0&35& 0&1&0&0&0&1&3&
4.40643&163.8&-30.1& 8.35225&285& 86& 16&43.7& 1.6&$  3.9\cdot 10^{-15}$&   6.0\\
 1684&99-308 21:21&1&2&116& 9& 7&10&12&15& 0&1&1& 8&31&1&0&0&0&1&3&
4.38865&163.9&-30.4& 8.32156& 31&234& 53& 5.0& 1.9&$  2.3\cdot 10^{-12}$&  10.5\\
 1685&99-315 08:11&3&3&189&20&26&17&10& 8& 6&0&1&44& 0&1&0&0&0&1&3&
4.36452&164.1&-30.9& 8.28016&137&297&-36&11.5& 1.9&$  1.4\cdot 10^{-11}$&  11.1\\
 1686&99-318 18:54&3&4&129&29&51&23& 8& 9& 5&0&1&46& 0&1&0&0&0&1&3&
4.35113&164.1&-31.2& 8.25732& 53&252& 34&10.4& 1.9&$  4.1\cdot 10^{-10}$&  10.5\\
 1687&99-323 02:34&0&1&134& 5& 9& 0& 7&15& 0&1&0&45&31&1&0&0&0&1&3&
4.33474&164.2&-31.5& 8.22950& 64&259& 25& 8.9& 4.3&$  2.9\cdot 10^{-13}$& 179.4\\
&&&&& &&&&& &&&&& &&&&& &&&&& &&&&& &\\                      \hline                                                      
 1688&99-326 06:21&0&1&182& 2& 2& 0& 9&15& 0&1&0& 0& 0&1&0&0&0&1&3&
4.32211&164.3&-31.7& 8.20815&132&297&-32&11.8&11.8&$  2.8\cdot 10^{-14}$&5858.3\\
 1689&99-333 04:30&0&1&249& 3& 6& 0& 8& 9& 9&0&0& 0& 0&1&0&0&0&1&3&
4.29512&164.4&-32.3& 8.16279&229& 60&-30&25.9& 1.6&$  4.1\cdot 10^{-15}$&   6.0\\
 1690&99-345 21:50&0&2& 26& 9&19& 0& 8& 6& 6&0&0&37& 0&1&0&0&0&1&3&
4.24406&164.7&-33.2& 8.07803&281& 92& 12&21.4& 1.9&$  1.2\cdot 10^{-13}$&  10.5\\
 1691&99-346 04:06&1&1&110& 6& 2& 5& 9&15& 0&1&1& 0& 0&1&0&0&0&1&3&
4.24305&164.7&-33.3& 8.07635& 39&241& 45&14.1& 1.9&$  3.4\cdot 10^{-14}$&  10.5\\
 1692&99-350 00:48&3&4&142&26&49&23& 8& 4& 5&0&1&46& 0&1&0&0&0&1&3&
4.22726&164.8&-33.6& 8.05043& 87&273&  5&10.4& 1.9&$  1.6\cdot 10^{-10}$&  10.5\\
\end{tabular*} \end{sidewaystable}                                                     
\clearpage                                                                      
\thispagestyle{empty}
\begin{sidewaystable}                                                                   
\tabcolsep1.2mm
\tiny \hspace{-2cm}
\begin{tabular*}{23cm} {%@{\extracolsep{\fill}}                                  
rrrrrrrrrrrrrrrcrrrrcrrrrrrrrrrrrrrrrrrrrr}                                     
 \mc{1}{c}{No.}&               \mc{1}{c}{IMP. DATE}&         \mc{1}{c}{C}&                 \mc{1}{c}{AR}&                
 \mc{1}{c}{S}&                 \mc{1}{c}{IA}&                \mc{1}{c}{EA}&                \mc{1}{c}{CA}&                
 \mc{1}{c}{IT}&                \mc{1}{c}{ET}&                \mc{1}{c}{E}&                 \mc{1}{c}{E}&                 
 \mc{1}{c}{I}&                 \mc{1}{c}{PA}&                \mc{1}{c}{P}&                 \mc{1}{c}{E}&                 
 \mc{1}{c}{I}&                 \mc{1}{c}{E}&                 \mc{1}{c}{C}&                 \mc{1}{c}{P}&                 
 \mc{1}{c}{HV}&                \mc{1}{c}{R}&                 \mc{1}{c}{LON}&               \mc{1}{c}{LAT}&               
 \mc{1}{c}{D$_{\rm Jup}$}&     \mc{1}{c}{ROT}&               \mc{1}{c}{S$_{\rm LON}$}&     \mc{1}{c}{S$_{\rm LAT}$}&     
 \mc{1}{c}{V}&                 \mc{1}{c}{VEF}&               \mc{1}{c}{M}&                 \mc{1}{c}{MEF}\\              
 &                             &                             \mc{1}{c}{L}&                 &                             
 \mc{1}{c}{E}&                 &                             &                             &                             
 &                             &                             \mc{1}{c}{I}&                 \mc{1}{c}{I}&                 
 \mc{1}{c}{I}&                 &                             \mc{1}{c}{E}&                 \mc{1}{c}{V}&                 
 \mc{1}{c}{C}&                 \mc{1}{c}{C}&                 \mc{1}{c}{C}&                 \mc{1}{c}{C}&                 
 &                             &                             &                             &                             
 &                             &                             &                             &                             
 &                             &                             &                             \\                            
 &                             &                             \mc{1}{c}{N}&                 &                             
 \mc{1}{c}{C}&                 &                             &                             &                             
 &                             &                             \mc{1}{c}{T}&                 \mc{1}{c}{C}&                 
 \mc{1}{c}{C}&                 &                             \mc{1}{c}{T}&                 \mc{1}{c}{D}&                 
 \mc{1}{c}{P}&                 \mc{1}{c}{P}&                 \mc{1}{c}{P}&                 \mc{1}{c}{P}&                 
 &                             &                             &                             &                             
 &                             &                             &                             &                             
 &                             &                             &                             \\                            
&&&&& &&&&& &&&&& &&&&& &&&&& &&&&& &\\                                         
\hline                                                                          
&&&&& &&&&& &&&&& &&&&& &&&&& &&&&& &\\                                         
 1693&99-355 22:26&0&2&222& 8&13& 0& 7& 7& 6&0&0&35& 0&1&0&0&0&1&3&
4.20309&164.9&-34.0& 8.01098&202& 33&-44&43.7& 1.6&$  4.7\cdot 10^{-15}$&   6.0\\
 1694&99-356 17:17&3&3&143&20&25& 8& 5& 6& 5&0&1&43& 0&1&0&0&0&1&3&
4.19999&164.9&-34.1& 8.00594& 91&276&  2&35.4& 1.6&$  2.8\cdot 10^{-13}$&   6.0\\
 1695&99-360 04:15&0&1& 22& 5& 2& 0& 7&15& 0&1&0& 0& 0&1&0&0&0&1&3&
4.18543&165.0&-34.4& 7.98235&283& 95& 13&34.1& 1.9&$  1.3\cdot 10^{-15}$&  10.5\\
\end{tabular*} \end{sidewaystable}                                                     

\clearpage

\section*{Figures}

\vskip -2cm

\nocite{landgraf2000b}

\nocite{gruen1997a} 
\nocite{witte1996} 

\end{document}